\documentclass[a4paper,fleqn,usenatbib]{mnras}
\usepackage[T1]{fontenc}
\usepackage{ae,aecompl}

\usepackage{lscape}
\usepackage{times,mathptm}
\usepackage{graphicx}	% Including figure files
\usepackage{amsmath}	% Advanced maths commands
\usepackage{amssymb}	% Extra maths symbols
\usepackage{hyperref}
\usepackage{tabularx}

\usepackage{pifont}% http://ctan.org/pkg/pifont
\def\apj{ApJ}
\def\apjl{ApJL}
\def\mnras{MNRAS}

\def\araa{ARAA}
\def\aap{A\&A}
\def\aapr{A\&AR}

\def\apjs{ApJS}

\title[Flat RCs of $z\sim 1$ SFGs]{Flat Rotation Curves of $z\sim 1$ Star-Forming Galaxies}

\author[G. Sharma et al.]{
Gauri Sharma,$^{1,2,3,5}$\thanks{E-mail: gsharma@sissa.it (GS)}
Paolo Salucci,$^{1,2,3}$
C. M. Harrison,$^{4}$
Glenn van de Ven,$^{5}$
Andrea Lapi$^{1,3}$
\vspace*{6pt}\\
$^{1}$SISSA International School for Advanced Studies, Via Bonomea 265, I-34136 Trieste, Italy\\
$^{2}$QGSKY, INFN-Sezione di Trieste, via Valerio 2, I-34127 Trieste, Italy\\
$^{3}$IFPU Institute for Fundamental Physics of the Universe, Via Beirut, 2, 34151 Trieste, Italy \\
$^{4}$School of Mathematics, Statistics and Physics, Newcastle University, Newcastle upon Tyne, NE1 7RU, UK\\
$^{5}$Department of Astrophysics, University of Vienna, T\"urkenschanzstrasse 17, 1180 Wien, Austria
}

\date{Accepted 2021 January 24. Received 2020 December 21; in original form 2020 April 30}

\pubyear{2020}

\begin{document}
\label{firstpage}
\pagerange{\pageref{firstpage}--\pageref{lastpage}}
\maketitle

\begin{abstract}
 We investigate the shape of the Rotation Curves (RCs) of $z\sim 1$ Star-Forming Galaxies (SFGs) and compare them with local SFGs. For this purpose, we have used $344$ galaxies from the K-band Multi-Object Spectrograph (KMOS) for Redshift One Spectroscopic Survey (KROSS). This sample covers the redshift range $0.57\leq z \leq 1.04$, the effective radii $0.69 \leq R_e \ \mathrm{\left[kpc \right]} \  \leq 7.76$, and the stellar masses $ 8.7 \leq \log\left(M_* \ \mathrm{\left[M_\odot \right]} \right) \leq 11.32$. Using $^{3D}$BAROLO, we extract the $H\alpha$ kinematic maps and corresponding RCs. The main advantage of $^{3D}$BAROLO is that it incorporates the beam smearing in the 3D observational space, which provide us with the intrinsic rotation velocity even in the low spatial resolution data. We have corrected the RCs for pressure support, which seems to be a more dominant effect than beam smearing in high-$z$ galaxies. Only a combination of the three techniques (3D-kinematic modelling + 3D-beam smearing correction + pressure gradient correction ) yields the intrinsic RC of an individual galaxy. Further, we present the co-added and binned RCs constructed out of 256 high-quality objects. We do not see any change in the shape of RCs with respect to the local SFGs. Moreover, we notice a significant evolution in the stellar-disk length ($R_D$) of the galaxies as a function of their circular velocity. Therefore, we conclude that the stellar disk of SFGs evolves over cosmic time (from $z\sim 1$) while the total mass stays constant (within $\sim 20 \ \mathrm{kpc}$). 
\end{abstract}


%%%%%%%%%%%%%%%%%%%%%%%%%%%%%%%%%%%%%%%%%%%%%%%%%%

%%%%%%%%%%%%%%%%% BODY OF PAPER %%%%%%%%%%%%%%%%%%
\section{Introduction}
\label{sec:Intro}
In the late 1980s, \citet{rubin1980} and \citet{bosma1981} published the most explicit observational evidence of non-Keplerian Rotation Curves (RCs) of spiral galaxies. These findings have made far-reaching changes in the field of Astronomy, Astrophysics as well as Cosmology, introducing an elusive component that astrophysicist dubbed as Dark Matter, thought to made of a dark particle necessarily beyond the standard model of elementary particles. Since then, Dark Matter (DM) became a building block of the current cosmological model so as the formation and evolution of all the structures in the Universe \citep{pad1993book, springel2005}. It contributes $\approx 24\%$ to the energy budget of the Universe \citep{freedman2003}, despite "no success" in the discovery of its particle nature.

In the local Universe, by studying the shape of the RCs, we established a fair understanding about the presence of DM and its contribution in the mass distribution \citep[e.g.,][and references therein]{PB2000, Sofue2001, PS2007, Courteau2015, PS2019}. The rotation curve studies not only constrain the mass budget but has also strengthened our understating of galaxy formation and evolution  \citep[e.g.,][]{reyes2011, read2016, karukes2017, lapi2018}. On the basis of observational evidence of the DM in the local Universe, and the development in the field of \textit{1)} high-resolution numerical simulations and \textit{2)} large galaxy surveys, those could identify the structures and substructures hosting galaxies and measure their spatial clustering. The current galaxy formation and evolution scenario suggest a theoretical account of DM halo, in which baryonic matter collapses to form the stars and a subsequent growth leads to the formation of a galaxy \citep[][references therein]{Wechsler2018}. 

In the last decade, advanced use of integral field units (IFUs) in galaxy surveys has opened the several possibilities of studying the spatially resolved kinematics and the dynamics of galaxies. For example, surveys with the Multi-Unit Spectroscopic Explorer (MUSE: \citealt{MUSE}), K-band Multi-Object Spectrograph (KMOS: \citealt{sharples_2014}), and the Spectrograph for INtegral Field Observations in the Near Infrared (SINFONI: \citealt{SINFONI}). Usually, the kinematics of the galaxies are derived using spatially-resolved emission line measurements (e.g., H$\alpha$, [OIII]) extracted from the IFU data. In particular in this work we use data from the KMOS Redshift One Spectroscopic Survey \citep[KROSS:][] {stott2016}, which uses the H$\alpha$ emission line to trace the galaxy kinematics.

\citet{Lang2017} and \citet{genzel2017}, used IFU data from KMOS and SINFONI to analyse the RCs of $0.6 \lesssim z \lesssim 2.4$ Star-Forming Galaxies (SFGs) and found a declining behaviour with increasing radius, in constrast to the RCs of local SFGs that are remarkably flat and rarely decline \citep[e.g.,][]{rubin1980, PS1996}. In brief, \citealt{Lang2017} studied the stacked normalized RCs of 101 SFGs at $0.6 \lesssim z \lesssim 2.2$ with stellar mass $9.3 \lesssim log(M_* \ \mathrm{[M_\odot]}) \lesssim 11.5$, where the normalization is performed at turn over radii, where $R_{turn} \sim 1.65R_e$. \citealt{genzel2017} studied the individual RCs of six massive ($log(M_* \ \mathrm{[M_\odot]})=10.6-11.1$) SFGs at redshift $0.9\leq z\leq 2.4 $. They showed the declining RCs in two cases: \textit{1)} when individual RCs are normalized at $R_{max}$ where the amplitude of rotation velocity is maximum and; \textit{2)} when binned averages of the six individual galaxies are normalized at the effective radii ($R_e$). In the end, both studies \citep{Lang2017, genzel2017} proposed that the declining behaviour of RCs can be explained by a combination of `high baryon fraction' and extensive `pressure support'. 

In comparison, \citet{AT2019a} studied the shape of RCs of $\approx 1500$ high-$z$ ($0.6\lesssim z \lesssim 2.2$) SFGs with stellar masses $8.5 \lesssim log(M_* \ \mathrm{[M_\odot]}) \lesssim 11.7 $. They used a similar stacking approach as \citet{Lang2017}, but normalized the RCs at three times the stellar disk scale length  (i.e., $3R_ {D} $, where $R_D = 0.59R_e$), without accounting the pressure support corrections. They found flat RCs, more like those seen in the local Universe. In the end, to explain the difference in their results to those presented in \citet{Lang2017}, \citet{AT2019a} concluded that the shape of stacked RCs depends on the choice of the normalization scale used in constructing the average RCs.

Differences in RC shapes may also arise due to different kinematic modelling approaches, different treatment of observational uncertainties (e.g., low resolution and small angular size lead to the beam smearing) and the underlying physical effects, e.g., pressure support/gradient \citep{val2007, read2016, SW2019}. In regards to the observational uncertainties, although IFUs brought remarkable progress in the field, due to the small angular size of the high-z galaxies, the attained spatial resolution is limited. Without Adaptive Optics (AO), an IFU achieves only $0.5\arcsec -1.0 \arcsec$ spatial resolution, whereas, a galaxy from $z\sim 1$ has a typical angular size of $2\arcsec - 3\arcsec$. The finite beam size causes the line emission to smear on the adjacent pixels. This effect is referred to as `Beam Smearing', which under estimates the rotation velocity and overestimates the velocity dispersion. The same beam smearing scenario happens in HI observations \citep{bosma1979, begeman1989} of local spiral galaxies. Although the previous IFU studies of high-z galaxies have applied beam-smearing corrections in different ways, these have usually been applied to the derived two-dimensional velocity maps or the one-dimensional RCs . An alternative approach is to apply dynamical models and beam-smearing corrections simultaneously directly to the 3D data cube.  For example, $^ {3D} $BAROLO \citep{ETD15, ETD16} (hereafter BBarolo) uses a tilted ring approach, which allows the reconstruction of intrinsic kinematics closest to the observations. Then the model is compared with data ring by ring in 3D-space, and, at the same time, beam smearing corrections are accounted for. This is the approach that we employ in this work and the details are discussed in Section~\ref{sec:Kmodelling}. 

In regards to varying physical galaxy properties affecting the RCs of galaxies, it is particularly important to consider that the interstellar medium (ISM) is turbulent in high-$z$ galaxies which could modify the kinematics of the galaxies \citep{Burkert2010, Glazebrook2013, Turner2017, HLJ17, Ubler2019, SW2019} and may result in different shapes of RCs. In fact, previous studies have shown that average gas velocity dispersion evolves with redshift as well as the disk fraction \citep{Kassin2007, Kassin2012, Wisnioski2015, Simons2017, Wisnioski2019} and dark matter content\footnote{A recent study by \citet{Genzel2020} shows that the dark matter versus baryonic matter contribution to the RCs may be different at different redshifts.}\citep[][and references therein]{Forster2020}. This could clearly impact upon the shape of the RCs that are derived from kinematic measurements of high-$z$ galaxies.

As mentioned earlier in the section, the kinematics of the galaxies are derived using the emission lines like H$_\alpha$ or [OIII]. These emission lines arise from the gaseous disk around the stars or the ISM. If the ISM is highly turbulent, then the emission also experiences a turbulence, i.e., radial force against gravity. This turbulence/force scales with the gas density and velocity dispersion. Since the density and velocity dispersion both decrease with increasing radius, this creates a pressure gradient, i.e., $F_{P} \propto -dP/dr$ (where $P \propto \rho \sigma^2$). The resulting radial force supports the disk and makes it rotate slower than the actual circular velocity, which might result in declining the RCs and potentially underestimate the dynamical masses \citep{val2007, Dalcanton_2010}. This effect is generally minimal in the local rotation-dominated SFGs but significant in the local dwarfs and early-type galaxies (e.g., \citealt{val2007, read2016, Anne2008}). Since, high-$z$ SFGs are gas dominated \citep[][references therein]{Glazebrook2013, Tacconi2018}, and the ISM is relatively turbulent \citep[][references therein]{Forster2020}. Therefore, it is essential to take into account the pressure gradient. In this work, we apply the `Pressure Gradient Correction' (PGC) on RCs, as mentioned in the Section~\ref{sec:ADC}.

The article is organized as follows: In Section~\ref{sec:data}, we describe the sample used in this work; Section~\ref{sec:modelling}, contains a brief discussion on the kinematic modelling using $^{3D}$Barolo code and Pressure Gradient Corrections; In the Section~\ref{sec:results} \& Section~\ref{sec:discussion}, we have discussed the main results, shown the shape of RCs, and their comparison with the locals RCs; Section~\ref{sec:summary} contains a summary of the work. In this work, we have assumed a flat $\Lambda$CDM cosmology with $\Omega_{m,0} =0.27$, $\Omega_{\Lambda,0}=0.73$ and $H_0=70 \ \mathrm{km \ s^{-1}}$.

%%%%%%%%%%%%%%%%%%%%%%%%%%%%%%%%%%%%%%%%%%%%%%%%%%%%%%%%%%%%%%%%%%%%%%%%
\section{DATA}
\label{sec:data}
KMOS-Redshift One Spectroscopic Survey (KROSS) was aimed to observe the $z\sim 1$ SFGs \citep{stott2016}. In this work, we have analysed a sub-sample of the publicly available KROSS data to determine the `intrinsic RCs' of high-$z$ rotation dominated SFGs (most likely disk-type galaxies). The minor and major details of observations and physical properties of the full sample can be found in \citet{stott2016} and other first and foremost papers by the KMOS team \citep[e.g.,][]{H17, HLJ17, AT2019a}. Nevertheless, in the section below, we have given a short overview of KROSS and our sample selection criteria.

		%%%%%%%%%%%%%%%%%%%%%%%%%%%%%%%%%%%%%%%%%%%%%%%%%%%%%%
\subsection{KMOS Observations}
\label{sec:kross}
KROSS is an Integrated Field Spectroscopic (IFS) survey using the KMOS instrument on ESO/VLT. The KMOS consists of 24 Integrated Field Units (IFUs); those can be placed within $7.2\arcmin$ diameter field. Each IFU covers the $2.8\arcsec \times 2.8\arcsec $ in size with $0.2\arcsec$ pixels. The targets for the survey are selected from extragalactic deep field covered by multi-wavelength photometric and spectroscopic data: 1)Extended Chandra Deep Field Survey (E-CDFS: \citealt{ECDFS1, ECDFS2}), 2)Cosmic Evolution Survey (COSMOS: \citealt{COSMOS}),  3)Ultra-Deep Survey (UKIDSS: \citealt{UKIDSS}), 4)SA22 field \citet{SA22}.Some of the targets were selected from CF-HiZELS survey \citep{CFHIZELS}. The targets were selected such that the $H_\alpha$ emission is shifted into J-band. The median redshift of parent sample (KROSS full sample) is $z=0.85^{+0.11}_{-0.04}$. The median J-band seeing of observations was $0.7\arcsec$, with $92$\% of the objects were observed during seeing $<1\arcsec$. Individual frames have exposure times of $600 \ sec$, and a chop to the sky was performed every two science frames. The data were reduced using ESOREX/SPARK pipeline \citep{davies2013}, and flux calibration is performed using standard stars which have been observed during the same night as science data. The end product of the process is 3D datacube consists of two spatial axes and one spectral axis of 2048 channels (e.g., 3D datacube = {\it f}($x,y,\lambda$)). These datacubes are capable of producing spectrum, the line and the continuum images and the moment maps (see: \citealt{stott2016}). Since mid-2019, this data is publicly available at KROSS-website\footnote{http://astro.dur.ac.uk/KROSS/data.html}.
		%%%%%%%%%%%%%%%%%%%%%%%%%%%%%%%%%%%%%%%%%%%%%%%%%%%%%%
\subsection{KROSS Sample Selection}
\label{sec:sample}
We are focusing on 586 KROSS galaxies studied by \citet[][hereafter H17] {H17}, we refer it as parent sample. We have selected 344 objects out of 586, on the basis of integrated $H\alpha$ flux cut ($F_{H\alpha}>2\times 10^ {-17} \  \mathrm{[erg \ s^ {-1} \ cm^ {-2}]} $) and inclination angle ($25^{\circ} \leq \theta_i \leq 75^{\circ} $). The chosen flux and inclination cuts ensure the sufficient signal-to-noise data (S/N) and reduces the impact of extinction
\footnote{In the highly inclined system ($\theta_i>75^{\circ}$) observed flux extinct due to extinction, which suppress the rotation velocity upto a few times $R_e$ \citep[see][]{Valotto2003}. On the other hand, in face-on galaxies ($\theta_i<25^{\circ}$) the rotation signal drops below the observational uncertainties. Therefore, to be conservative we down-select the sample for $25^{\circ} \leq \theta_i \leq 75^{\circ} $.} during the kinematic modelling procedure (see Section~\ref{sec:Kmodelling}).  The intrinsic characteristic of the selected sample (referred to as `analysed sample') is the following (given with respect to TableA1 of H17): 1)AGN-flag is zero i.e., no evidence for an AGN contribution to the $H_\alpha$ emission-line profile; 2)H17 Quality-flag 1, 2, and 3,  i.e., only $H_\alpha$-emission line detected objects ($S/N>3$). We adopted the values of effective radii ($R_e$), photometric position angle ($PA$), photometric inclination angle ($\theta_{i}$), absolute H-band magnitude ($M_H$), K-band AB magnitude ($K_{AB}$), z-band AB magnitude ($z_{AB}$), $H_\alpha$ luminosity ($L_{H\alpha}$), $H_\alpha$ flux ($F_{H\alpha}$), $H_\alpha$ star-formation rate ($SFR_{H\alpha}$) and redshift ($z$). For the details of adopted quantities we refer the reader to \citet{H17} and \citet{stott2016}, while next we briefly discuss some of the requisite quantities.

The position angle ($PA$) and inclination angle ($\theta_{i}$) are estimated by fitting a two-dimensional Gaussian model to the broadband images. \citet{H17} compared their $PA$ and $\theta_{i}$ with \citet{vanderWel2012} which fits S\'ersic models to the {\em HST} near-infrared images using {\sc galfit} that incorporates PSF modelling. Their calculations were in agreement with the {\sc galfit} results and those derived using a two-dimensional Gaussian fitting method. Moreover, $PA$ and $\theta_{i}$ for COSMOS targets with $I$-band images were cross-checked with \citet{Tasca2009} who derived the $PA$ and $\theta_{i}$  using the axis ratios. 

The effective radii ($R_e$) is measured from the broadband images by deconvolving the PSF and semi-major axis of the aperture, which contains half of the total flux. Since the broadband images are observed in $I, z^\prime, H$ and $K$ bands (depends on the surveys goal and instrument facility) therefore, the targets where the images are in $I$, and $z^\prime$ band, a systematic correction factor of 1.1 is applied in $R_e$ to account the colour gradient. The simple approach of colour correction applied because HST images were not available for all the targets (for details see \citealt{H17}). 

To obtain the galaxy integrated $H_\alpha$ luminosity ($L_{H\alpha}$), \citet{H17} first grade the $H_\alpha$ sources on the basis of signal-to-noise (S/N: average over two times derived velocity FWHM of $H_\alpha$ line). If the S/N $\leq 3$, then sources are discarded. Second, the emission line width is corrected for instrumental dispersion, which is measured from unblended skylines near the observed wavelength of the $H_\alpha$ emission. Then, the $H_\alpha$ flux was measured using $2.5 \arcsec $ aperture (with an uncertainty of $30\%$) and hence the integrated $H_\alpha$ luminosity\footnote{Notice, we do not account for extinction due to lack of required data (e.g., Balmer line ratios) to measure the extinction. However, these luminosities are not used for the bulk of our analyses.} obtained.

The Stellar masses are derived using the {\sc Le Phare} \citep{Arnouts1999, Ilbert2006} Spectral Energy Distribution (SED) fitting tool. The {\sc Le Phare} compares the suits of modelled SED of object from observed SED. Where observed SED of our sample is derived from optical \& NIR photometric bands (U, B, V, R, I, J, H, and K), in some cases we have also used the IRAC mid-infrared bands. In modelling, the stellar population synthesis model is derived from \citet{Bruzual2003} and stellar masses are calculated using \citet{Chabrier_2003} Initial Mass Function (IMF). The {\sc Le Phare} routine fits the extinction, metallicity, age, star-formation, and stellar masses and allows for a single burst, exponential decline and constant star formation histories. For details of stellar mass computation we refer the reader to \citet{AT2019}.

We remark, the parent sample is selected in such a way that it does not preferentially contain galaxies in a merging or interacting state,and include good representatives of main sequence SFGs at $z\sim 1$ \citep[see][and references therein]{stott2016}. To reaffirm, we present the distributions of the physical quantities (namely,  $F_{H\alpha}$, $\theta_i$, $z$, $R_e$, $M_*$, and $SFR_{H\alpha}$) of the parent and analysed sample, which are shown in Figure~\ref{fig:KROSS-hist}. In the Appendix~\ref{fig:extrahist}, we additionally provide the distribution of the $L_{H\alpha}$ and absolute H-band magnitude. In short, our analysed sample covers the following range of the $H_\alpha$ flux: $-16.70\leq \log(F_{H\alpha} \ \mathrm{[erg \ s^{-1} \ cm^{-2}]})\leq -15.26$, the inclination angle: $25^{\circ} \leq \theta_i \leq 75^{\circ}$, the redshift: $0.75\leq z \leq1.04$, the effective radii: $-0.16\leq \log(R_e \ \mathrm{[kpc]}) \leq 0.89$, the stellar mass: $8.79 \leq \log(M_* \ \mathrm{[M_\odot]}) \leq 11.32$, and the star formation rate:$0.15 \leq \log(SFR_{H\alpha} \ \mathrm{[M_\odot \ yr^{-1}]}) \leq 1.76$. The final analysed sample selected for this study is representative of the parent sample from \citet{H17} and, therefore, is representative of main sequence, star-forming galaxies at this redshift.

\begin{figure*}
	\begin{center}
		\includegraphics[angle=0,height=3.0truecm,width=5.0truecm]{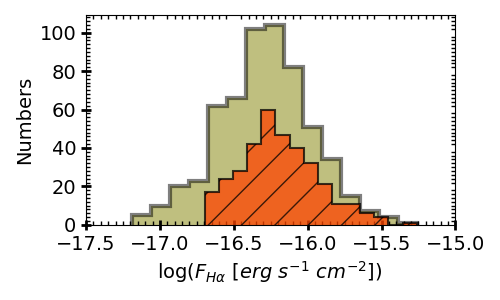}
\includegraphics[angle=0,height=3.0truecm,width=5.0truecm]{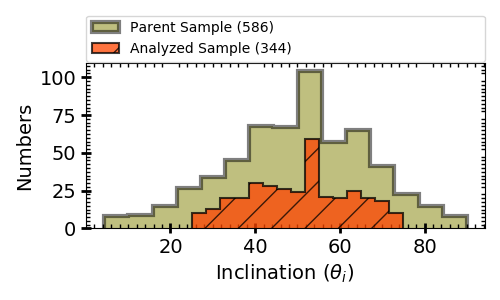}
\includegraphics[angle=0,height=3.0truecm,width=5.0truecm]{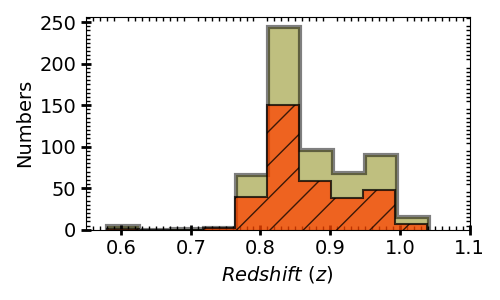}
\includegraphics[angle=0,height=3.0truecm,width=5.0truecm]{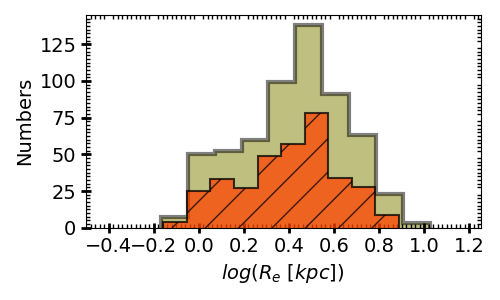}
\includegraphics[angle=0,height=3.0truecm,width=5.0truecm]{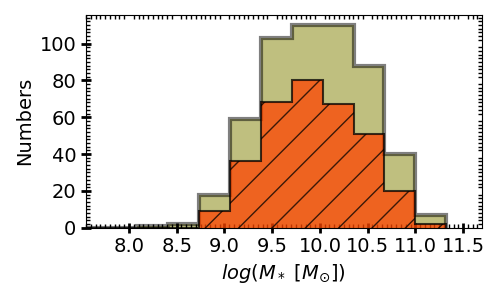}
\includegraphics[angle=0,height=3.0truecm,width=5.0truecm]{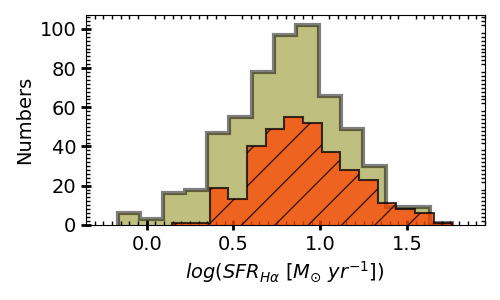}

		\caption{The distributions of physical quantities of parent and analysed sample. The color code is same in all the panels and given as follows, the parent KROSS sample (given by \citealt{H17}) shown by the green histograms and the analysed sample by orange hatched histogram. The {\em upper left \& middle panel} represents the $F_{H\alpha}$ and $\theta_i$ distribution, which verifies our sample selection cuts. {\em Upper right panel} shows the redshift ($z$) distribution, this justify that we are working with $z\sim 1$ galaxies. {\em Lower left Panel} demonstrate the size distribution of our analysed sample. The {\em lower middle \& right panel} shows the distribution of stellar masses and $H_\alpha$ based star-formation rate (SFR), these distributions are informative of main sequence at $z\sim 1$.}
		\label{fig:KROSS-hist}
	\end{center}
\end{figure*} 

%%%%%%%%%%%%%%%%%%%%%%%%%%%%%%%%%%%%%%%%%%%%%%%%%%%%%%%%%%%%%%%%%%%%%%%%
\section{METHODS}
\label{sec:modelling}
To obtain the intrinsic shape of the RCs, we follow the 3D-forward modelling and a relatively rigorous approach is considered for handling the observational and the physical uncertainties. Under 3D-forward modelling, a simulated datacube is populated for given initial conditions and then compared with the observed datacube. We keep on populating/reconstructing the simulated datacube by changing the initial guess until convergence between data and model occur. This yields the  PV-diagrams and moment maps. For implementing such 3D-forward kinematic modelling of datacubes, we have used $^{3D}$BAROLO code \citep{ETD15}, it is discussed in the section below. In the end, we implement the pressure support correction on  $^{3D}$BAROLO generated RCs, which is discussed in the Section~\ref{sec:ADC}.

		%%%%%%%%%%%%%%%%%%%%%%%%%%%%%%%%%%%%%%%%%%%%%%%%%%%%%%
\subsection{KINEMATIC MODELLING WITH $^{3D}$BAROLO}
\label{sec:Kmodelling}
We have modelled the kinematics of our sample using the $^{3D}$BAROLO code \citep{ETD15}.  The main advantage of modelling the datacube with $^{3D}$BAROLO (BBarolo): (1) it allows us to reconstruct the intrinsic kinematics in 6-domain (three spatial and three velocity components)  for given initial conditions; (2) the 3D projected modelled datacube is compared to the observed datacube in 3D-space; (3) it simultaneously incorporates the instrumental and the observational uncertainties (e.g., spectral-smearing\footnote{line spread function (LSF) which corresponds to spectral broadening} and beam-smearing\footnote{the point spread function (PSF) which determines the spatial resolution}) in 3D-space. For details, we refer the reader to \citet[][]{ETD15} and \citet{ETD16}. This 3-fold approach of deriving kinematics is designed to overcome the observational and instrumental effects and hence allows us to stay close to the realistic conditions of the galaxy. Therefore, it gives us somewhat improved results than the 2D-approach of kinematic modelling on the datacubes, specifically, in the case of small angular sizes and moderate S/N of high-$z$ galaxies (see \citealt{ETD16}). In the section below, we have discussed the BBarolo's underlying assumptions, its initial requirements for performing the kinematic modelling, and the very first results/tests on a large dataset.
		%%%%%%%%%%%%%%%%%%%%%%%%%%%%%%%%%%%%%%%%%%%%%%%%%%%%%%
\subsubsection{Basic assumption under $^{3D}$BAROLO}
\label{sec:3Bassumption}
BBarolo is based on the "tilted ring model," i.e., the motion of the gas and stars are assumed to be in the circular orbits. It does not assume any functional evolution of the kinematic quantities (e,g., $v_{rot}(R) \propto \arctan(R)$). Therefore, free parameters in BBarolo are not forced to follow any parametric form, rather estimated in the annuli of increasing distance from the galaxy centre without making any assumption on their evolution with the radius. However, BBarolo uses the radial binning for the velocity measurement, because in the 'tilted ring model', a galaxy is divided into several rings and parameters are calculated within each ring. The number of pixels used per bin depends on the choice of NRADII\footnote{Number of rings used in fitting the galaxy} and RADSEP\footnote{Separation between rings in arcsec} in the fitting. Therefore, each position-velocity diagrams contains $\sim 3-6$ rotation velocity measurements (and similarly for the velocity dispersion). The errors on velocity measurement per radial bin (inside BBarolo) are estimated using Monte Carlo sampling. 

A non-parametric approach of calculating kinematic parameters, makes BBarolo robust and reliable to use, and this is one of the reasons we are using it for kinematic modelling. There are other high-$z$ 3D-kinematic modelling codes, e.g., \texttt{GalPak 3D} \citep[][]{Bouch2015}, which has been successfully tested on $z\sim 0.5$ galaxies observed from the Multi-Unit Spectroscopic Explorer (MUSE) but it follows the parametric approach and \texttt{BLOBBY3D} \citep[][]{BLOBBY3D}, which has been tested on $20$ local star-forming galaxies from the SAMI Galaxy Survey. It is a useful tool to study the gas dynamics of high-$z$ low-resolution data. However, it comes with a long list of free parameters, which one may not know without a detailed study of the system and will be particularly degenerate for lower S/N data available for high redshift systems.

 		%%%%%%%%%%%%%%%%%%%%%%%%%%%%%%%%%%%%%%%%%%%%%%%%%%%%%%
\subsubsection{Initial requirements of $^{3D}$BAROLO }
\label{sec:InitialParams}
The kinematic modelling with BBarolo requires three geometrical parameters, i.e. the coordinates of galactic centre in the datacube ($x_c,y_c$), the inclination angle ($\theta_i$), the position angle ($PA$) and three kinematic parameters, i.e., the redshift ($z$), the rotation velocity ($v_{rot}$) and the velocity dispersion of ionized gas ($\sigma_{H\alpha}$). In our modelling, we fix the geometrical parameters and redshift (with an exception for $PA$, discussed below) and leave the two kinematic parameters free. Notice, ($x_c,y_c$) are the photometric galactic centre positions adopted from H17. BBarolo comes with several useful features particularly necessary/useful for high-$z$ low S/N data (see BBarolo documentation\footnote{https://bbarolo.readthedocs.io/en/latest/}). We are using \texttt{3DFIT TASK} for performing the kinematic modelling. First, BBarolo produces the mock observations on the basis of given initial conditions in the 3D observational space ($x,y,\lambda$), where ($x,y$) stands for the spatial axes and $\lambda$ is spectral axis coordinate. These models are then fitted to the observed datacube in the same 3D-space accounting for the beam smearing simultaneously. A successful run of  BBarolo delivers the beam smearing corrected moment maps, the stellar surface brightness profile, the rotation curve (RC), and the dispersion curve (DC) along with the kinematic models. Notice, RCs/PV-diagrams are not derived from the velocity maps, instead they are  calculated directly from datacubes by minimizing $V_{LOS} = V_{rot} \sin \theta_i$ (of model and data). BBarolo is well tested on the local systems \citep[e.g.,][]{ETD15, Korsaga2019} as well as on high-$z$ galaxies including the KMOS data \citep[see:][]{ETD16, Loiacono2019}.

The position angle ($PA$) is usually fixed to the photometric $PA$ ($PA_{phot}$) adopted from the H17 catalog, but for $\sim 44$\% of objects, $PA_{phot}$ doesn't allow to extract position-velocity (PV) diagram. This might be a consequence of misaligned morphological and kinematic major-axis. We know the fact that photometric and kinematic $PAs$ are not necessarily the same. That is why for these $44\%$ objects, $PA$ is kept free and estimated from the BBarolo kinematic modelling. A short discussion and an example is shown in Appendix~\ref{sec:kin-phot-barolo} \& Figure~\ref{fig:PA_phot_kin_model}. In the end, we also present a quantitative measure of misaligned $PAs$, see Appendix~\ref{sec:misalignment}.

 		%%%%%%%%%%%%%%%%%%%%%%%%%%%%%%%%%%%%%%%%%%%%%%%%%%%%%%
\subsubsection{Limitations in $^{3D}$BAROLO}
\label{sec:limBarolo}
To obtain a good fit BBarolo requires a mask to identify the true emission region and to ignores the noisy pixels. For this purpose, we use the in-built \texttt{MASK} task with an input of either \texttt{SEARCH} or \texttt{NEGATIVE} masking. In particular,  the \texttt{SEARCH} mask uses the source finder algorithm \texttt{DUCHAMP} \citep{DUCHAMP} and hence builds a mask on the identified emission regions based on $3D$ reconstructed sources. This mask works very well on high S/N (clean) data if noise is Gaussian distributed among the channels. However, when the noise does not follow a Gaussian distribution, then the \texttt{SEARCH} task is unable to find the source. In this situation, we have used the \texttt{NEGATIVE} mask. It computes the noise statistics ($\sigma_{noise}$) channel by channel, using only pixels with negative values, then build the mask in regions of  $flux>  (1.5-2.5) \ \sigma_{noise}$. Notice, noise in negative pixels are often more Gaussian-distributed than positive pixels, especially in low S/N KMOS data, and hence returns a better estimate of noise properties.

Notice that fully exploiting the $3^{rd}$ (velocity) dimension is still a caveat in the 2D/3D kinematic modelling of high-$z$ galaxies. However, we suggest that 3D-forward modelling has a potential to provide better results than applying models to fit to 2D or 1D projections of the datacube, because it uses the full information available inside the datacube. We remark \citep[see also][]{ETD15} BBarolo corrects well for beam smearing whenever a galaxy is resolved with at least 2-3 resolution elements across and has a $S/N\gtrsim 3$. The relative errors are within $20$\% for the rotation curve and within a factor 2 for the velocity dispersion, in the worst cases.

The errors on the data are estimated during the fitting procedure, the algorithm weights pixels based on their S/N, i.e. a pixel with a high S/N is considered more reliable and given more weight than a pixel with low S/N. The noise level is calculated directly from emission-free regions of the datacube using robust statistics (median and absolute deviation from the median). Notice, BBarolo's current error-estimation algorithm does not account for intrinsic uncertainties in the data. However, we minimised the impact of these effects by a careful selection/analysis of sub-sample (see Section~\ref{sec:sample} \& \ref{sec:Bresults}).

Finally let us remark, extinction is complex to account in the 3D modelling therefore, it is not accounted in BBarolo. However, extinction is important in inner region of galaxies, where high-z galaxies are not resolved. At least in our sample, we do not have resolution within $\sim 2 \ \mathrm{kpc}$, where extinction dominates. In particular we do not draw any conclusions about the inner parts of the rotation curves (see Section~\ref{sec:CRCs}). Nevertheless, BBarolo's performance has been extensively tested on optical/NIR data (e.g., KROSS, KMOS$^{3D}$ \citealt{ETD15, ETD16, Korsaga2019,Loiacono2019}), where it delivers remarkable results. Therefore, we expect BBarolo to perform well on our sample too; however, results are critically analysed and discarded whenever required (see Section~\ref{sec:Bresults}).

 		%%%%%%%%%%%%%%%%%%%%%%%%%%%%%%%%%%%%%%%%%%%%%%%%%%%%%%
\subsubsection{Results from $^{3D}$BAROLO}
\label{sec:Bresults}
The analysed sample was selected on the basis of the $H_\alpha$ emission line flux cut ($F_{H\alpha}>2 \times 10^{-17} \ \mathrm{[erg \ s^{-1}\ cm^{-2}]}$), the inclination angle cut ($25^{\circ} \leq \theta_i \leq 75^{\circ}$), and the S/N ($>3$) of the $H_\alpha$ detection. After executing BBarolo, we have visually inspected the BBarolo-outputs for quality assessment. We noticed that the best quality outputs correspond to objects with the inclination $45^{\circ} \leq \theta_i \leq 75^{\circ}$ where kinematics is derived using the \texttt{SEARCH} mask. On the other hand, worse quality is often spotted for relatively low inclination ($\theta_i < 45^{\circ}$) along with use of the \texttt{NEGATIVE} mask. Moreover, as per the limitations of BBarolo, objects with low S/N and $\theta_i<45^{\circ}$ might be being considered as background galaxy emission, which may lead to over/underestimated rotation velocity or  velocity dispersion. Therefore, to classify the quality of the galaxies, we have taken into account the S/N per pixel in the masked region, limitations of BBarolo and the participation of the mask (which gives us a direct indication of the noise level). Hence, we have assigned the quality of BBarolo outputs as the following:
\begin{enumerate}
\item \textbf{Quality-1}: $45^{\circ}< \theta_i \leq 75^{\circ}$ with \texttt{SEARCH} mask and $S/N>3$ in masked region.
\item \textbf{Quality-2}: $30^{\circ} \leq \theta_i \leq 75^{\circ}$ with \texttt{SEARCH} or \texttt{NEGATIVE} mask and $S/N\geq 3$.
\item \textbf{Quality-3}: the remaining objects. 
\end{enumerate}

\begin{figure}
	\begin{center}
		\includegraphics[width=\columnwidth]{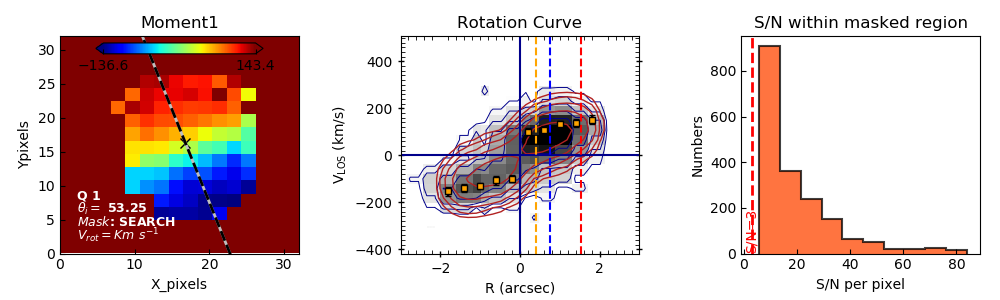}
		\includegraphics[width=\columnwidth]{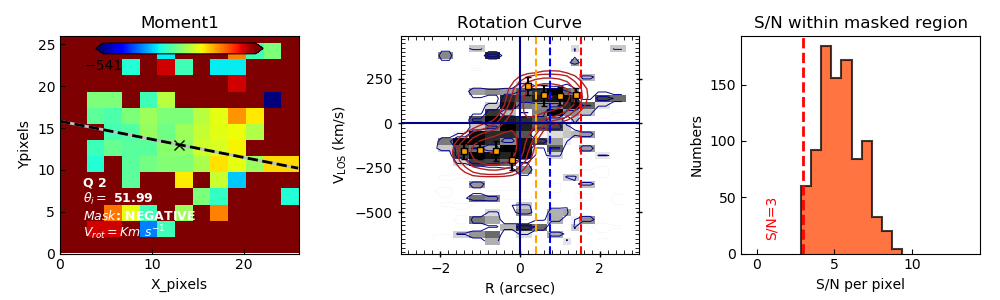}
		\includegraphics[width=\columnwidth]{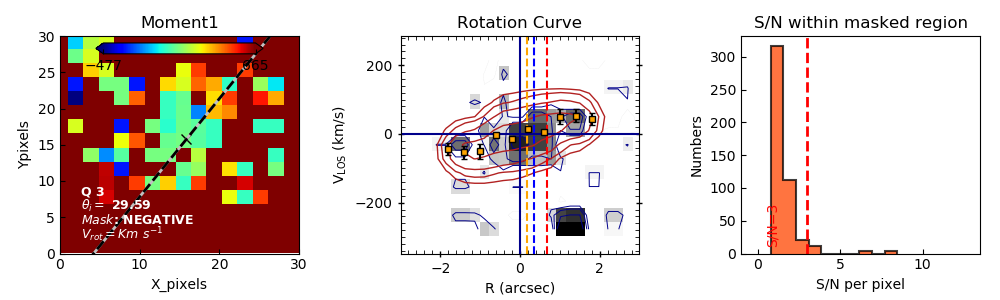}	
		\caption{A visual representation of the quality assessment of BBarolo outputs. The Upper, middle and bottom row represents the Quality-1, Quality-2 \& Quality-3 galaxies respectively. {\em COL 1:} First moment map, black-grey dashed line is showing the position angle, and the black cross shows the galactic centre ($x^p_c, y^p_c$). {\em COL 2:} Rotation curve, the black shaded area with blue contour shows the data while the red contour refers to the model and the orange squares with error bars are the best-fit velocity measurements. The yellow, blue and red vertical dashed lines are representing the effective radius ($R_e$), optical radius ($R_{opt}=1.89 \ R_e$), twice optical radius ($R_{out}$) respectively. {\em COL 3:} Distribution of S/N in masked region of datacube. The vertical red dashed line shows the $S/N=3$.}
		\label{fig:Q-gal}
	\end{center}
\end{figure}

An example of a visual representation of the Quality-1, 2, \& 3 objects is shown in Figure~\ref{fig:Q-gal}. In total, we have 120 Quality-1, 194 Quality-2 and 30 Quality-3 objects. The Quality-3 galaxies are discarded from the rotation curve analysis. We would remark, \citet{ETD15} have tested the performance of BBarolo on simulate data in terms of S/N, spatial/spectral resolution, and galaxy properties. They have shown that the code works well when most of the emission has $S/N \gtrsim 3$ and for inclination $30^{\circ} \leq \theta_i \leq 75^{\circ}$. All of our Quality-1 \& Quality-2 objects (Q12 sample) abide these criteria, except nine galaxies those inclined between $25^{\circ} <\theta_i < 30^{\circ}$ but contain $S/N>3$ per pixel. In the external appendix, we attach the plots of S/N per pixel of the analysed sample, and inclination is given in the catalog released with this paper.

\begin{figure*}
	\begin{centering}
		\includegraphics[angle=0,height=3.0truecm,width=18truecm]{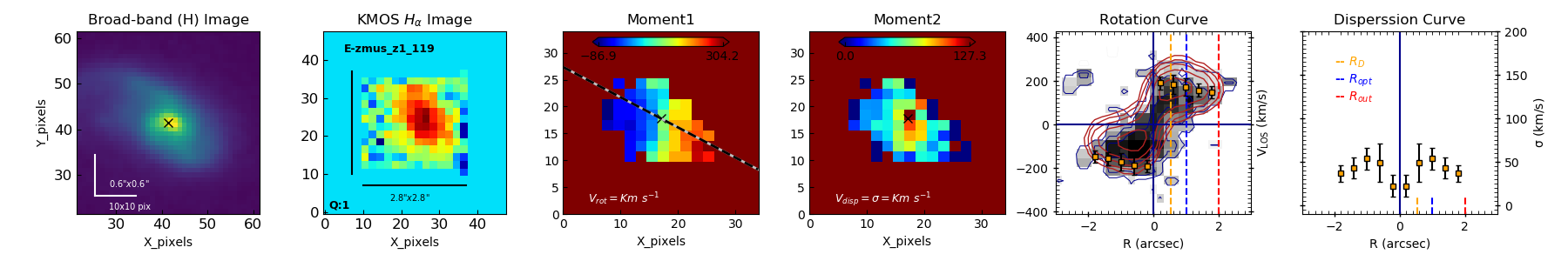}
		
		\includegraphics[angle=0,height=3.0truecm,width=18truecm]{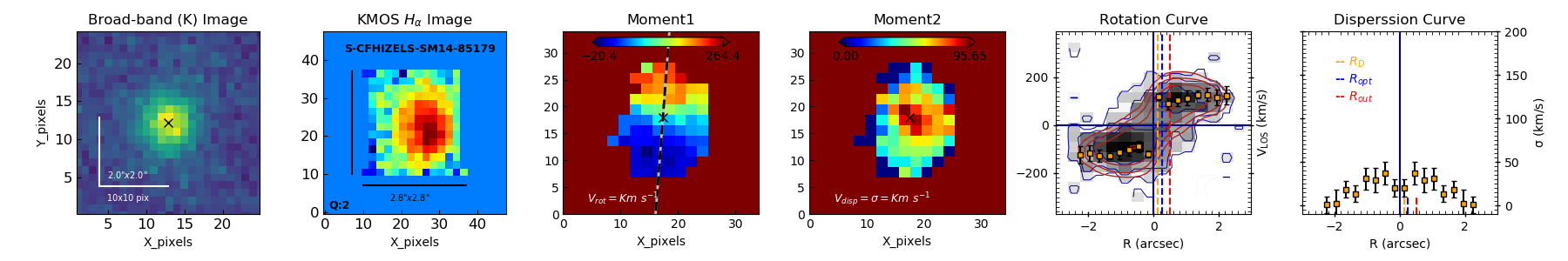}
		
		\includegraphics[angle=0,height=3.0truecm,width=18truecm]{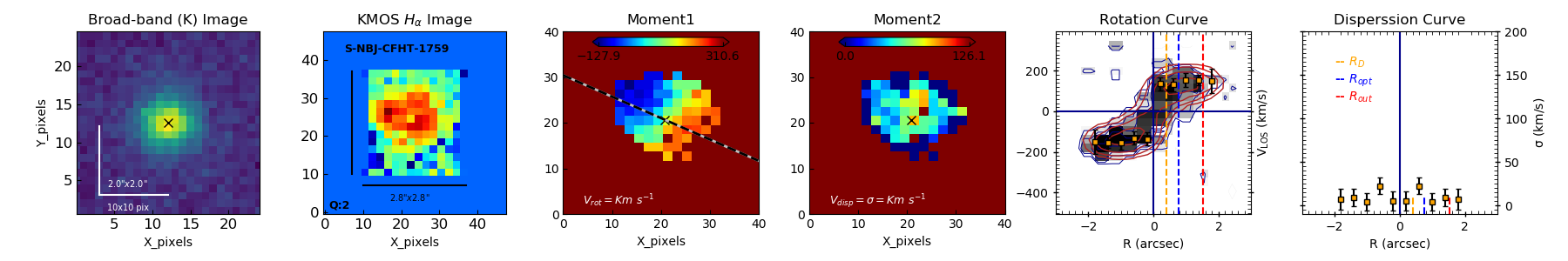}
		
		\includegraphics[angle=0,height=3.0truecm,width=18truecm]{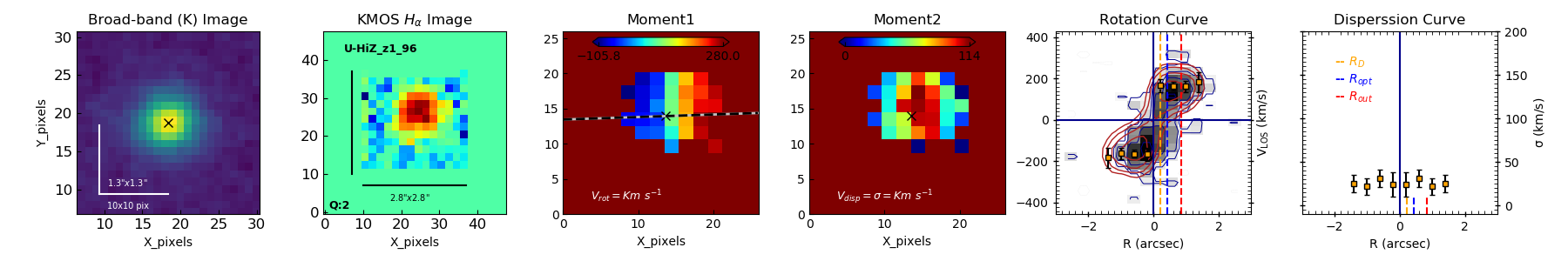}
		
		\includegraphics[angle=0,height=3.0truecm,width=18truecm]{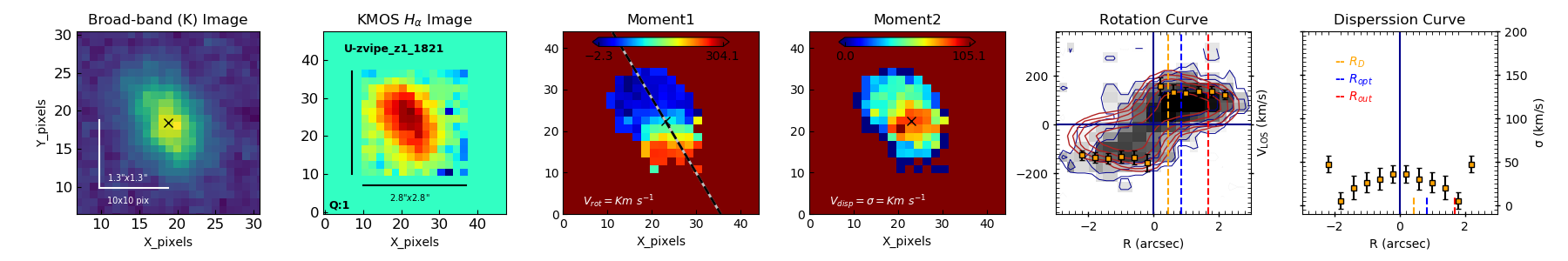}
		
		\includegraphics[angle=0,height=3.0truecm,width=18truecm]{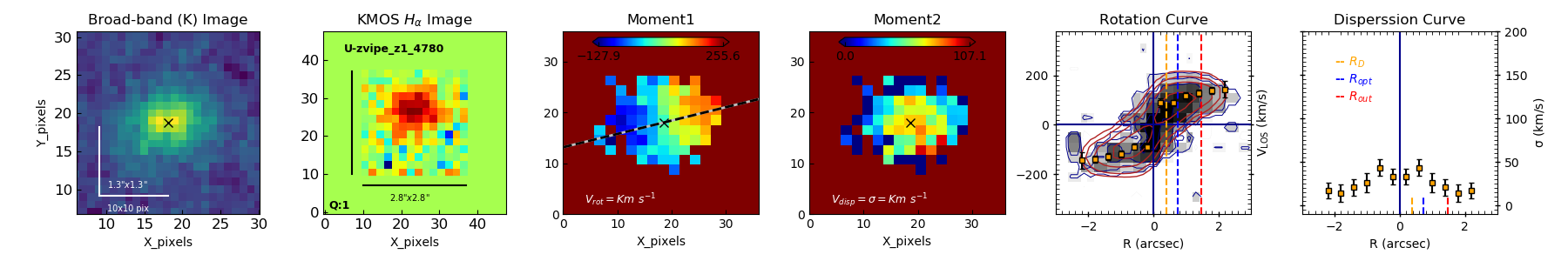}
		
		\includegraphics[angle=0,height=3.0truecm,width=18truecm]{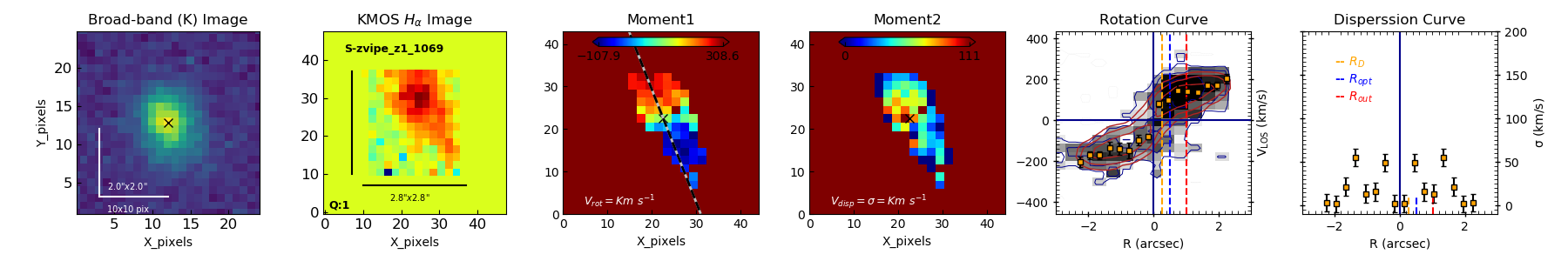}
		
		\caption{The outputs of the kinematic modelling using $^{3D}$BAROLO. {\em COL 1:} Broad-band image, where black cross shows the central coordinates of the photometric image. The horizontal and the vertical white lines in the lower-left corner are showing the $10\times10 \ pixel$ size in arcsec (an estimator of image-size). {\em COL 2:} Integrated $H_\alpha$-image from datacube, the size of the image is shown by the black horizontal and the vertical lines. The name of the galaxy is shown in the upper-left corner, and quality is mentioned in the lower-left corner. {\em COL 3-4:} the first and second-moment map, the black-grey dashed line is showing the position angle, and the black cross represents the galactic centre. {\em COL 5:} Rotation curve, the black shaded area with blue contour represents the data while the red contour shows the model and the orange squares with error bars are the best-fit rotation velocity measurements. {\em COL 6:} the best-fit velocity dispersion. The yellow, blue ,and the red vertical dashed lines are representing the effective radius ($R_e$), the optical radius ($R_{opt}=1.89 \ R_e$), twice the optical radius ($R_{out}=R_{2opt}=3.78 \ R_e$) respectively. Notice that the first data point in RCs and DCs is within resolution limit, therefore excluded from the analysis.}
		\label{fig:barolo-output}	
	\end{centering}
\end{figure*}

In Figure~\ref{fig:barolo-output}, we have shown a few examples of BBarolo outputs. From left to right, the broad-band high-resolution image, the $H_\alpha$-image, the first-moment (rotation velocity) map, the second-moment (dispersion velocity) map, the rotation curve, and the dispersion curve. In detail, {\em COL 1:} the broad-band image is constructed from the ground and the space-based photometric observational data (discussed in Section~\ref{sec:kross}). The central photometric coordinate (galactic centre: ($x^p_c,y^p_c$)) of the object is shown by a black cross, which is calculated by fitting the 2D Gaussian to the 2D distribution of the data. The size of each image can be inferred in terms of $10\times10 \ pixel$ size, converted to arcsec (displayed on the bottom left of the image). The size of the broad-band images varies from image-to-image due to the multi-wavelength data and different photometric surveys. {\em COL 2:} the integrated $H_\alpha$-image constructed from the KMOS datacube. The size of the image is displayed in arcsec by drawing the horizontal and the vertical black line. The name of the galaxy is displayed in the upper-left corner, and the quality displayed in the lower-left corner. {\em COL 3,4:} Moment-1 map and Moment-2 map, these maps are the output of the BBarolo kinematic modelling. The black-grey dashed line shows the position angle of the image. The black cross represents the galactic centre positions adopted from the work of \citet{H17}. {\em COL 5,6:} the rotation curve (RC) \& the dispersion curve (DC), constructed after a comparison of the data and the model in 3D-space (an output of BBarolo). In the rotation curve, the red contour is the model and the black shaded area with the blue contour represents the data. The orange squares with error bars are the best-fit rotation velocity (referred to as `RC data') and velocity dispersion. The yellow, blue and red vertical dashed lines are representing the effective radius ($R_e$), the optical radius ($R_{opt}=1.89 \ R_e$), and twice the optical radius ($R_{out}=R_{2opt}=3.78 \ R_e$) respectively. The size of the $H_\alpha$-image is always $2.8\arcsec \times 2.8\arcsec$, so are the spatial length of the moment maps, rotation and the dispersion curve. In some cases, even though $R_e$ is $1.6 \ kpc$ (i.e., $\ 0.2 \ arcsec$), RCs are extended more than $16 \ kpc$ (i.e., $\ 2.0 \ arcsec$) due to the fact that $H_\alpha$-emission can trace the light up to large radius in comparison to broad-band filters. A full version of Figure~\ref{fig:barolo-output} is attached in the external appendix. Notice, BBarolo estimates the errors using Monte Carlo sampling, which are plotted on RCs and DCs in Figure~\ref{fig:barolo-output}. In further analysis we consider symmetric errors on RC/DC data, since parameter space is very much Gaussian distributed. However, to be precise, we take root mean square of upper and lower bounds of BBarolo estimated errors.

 		%%%%%%%%%%%%%%%%%%%%%%%%%%%%%%%%%%%%%%%%%%%%%%%%%%%%%%
\subsection{Pressure Gradient Correction}
\label{sec:ADC}
A significant amount of the BBarolo generated RCs show a strong asymmetry and rapid fall in the inner region as well as in the outskirts of the galaxy, see left panel of Figure~\ref{fig:RC-ADCRCs}. Such a rise and fall could be either due to the low dark matter fraction or, it could be an impact of pressure support (e.g., \citealt{genzel2017}).  The latter is observed in local dwarfs and early-type galaxies (e.g., \citealt{val2007, Anne2008, read2016}), which noticeably suppresses the rotation velocity of the gas. In short, if the ISM is highly turbulent (like in high-$z$ galaxies \citealt{Burkert2010, Turner2017, HLJ17, Ubler2019, SW2019}), then the pressure gradient induces a force against gravity, which supports the disk against gravity and keeps it in kinematic equilibrium. This force is negligible in local rotation-dominated systems, but the same is not valid for high-$z$ galaxies. Mainly, in the case of the dynamical mass modelling of high-$z$ galaxies, it is very crucial to disentangle the pressure support; otherwise, one might lead to wrong estimates of baryonic and dark matter components. Therefore, to correct the azimuthal velocities for the pressure support, we follow \citet{Anne2008} by adopting their following Pressure Gradient Corrections (PGC):
\begin{equation}
\label{eq:Vadc0}
V_c^2 = V_{\phi}^2 + \sigma_R^2 \Big[- \frac{\partial \ ln \ \rho \ \sigma^2_R}{\partial \ ln \ R} + (1-\frac{\sigma^2_\phi}{\sigma^2_R})   \Big] 
\end{equation}
where $V_{\phi}$ is the inclination-corrected\footnote{Inclination correction is required, to go from observed to intrinsic rotation velocity as well as to convert the observed velocity dispersion into the intrinsic radial dispersion in the general case of an anisotropic velocity distribution (see eq.~A13 of \citealt{Anne2008}).} rotation velocity, $\rho$ is the density of gas,  $\sigma_R$ and $\sigma_\phi$ are the intrinsic radial and vertical velocity dispersion respectively. Under the common assumption of a constant disk scale height the slope of the intrinsic 3D-density ($\rho$) and 2D-surface density are the same; therefore, $\rho $ can be replaced with $\Sigma$, where $\Sigma$ is 2D-density\footnote{In fact, $\rho = \kappa \  \Sigma$, where $\kappa$ is constant with radius.} proportional to $H_\alpha$ mass surface density. From the kinematic modelling of the datacubes (discussed in Section~\ref{sec:Kmodelling}), we have required information about $\Sigma, \sigma_R$ and $V_\phi$ to employ into the PGC. Let us remark, all the quantities (namely: $V_\phi$, $\sigma_R$, and $\Sigma$) are function of radius ($R$)\footnote{i.e., $V_\phi = V_\phi (R)$, $\sigma_R = \sigma_R (R)$, $\Sigma = \Sigma(R)$, and $\alpha = \alpha(R)$}, and they are derived from $H_\alpha$ datacubes.

In Equation~\ref{eq:Vadc0}, the second term ($- \partial ln \ \rho \sigma_R^2 / \partial ln \ R$) gives the pressure gradient and the third term ($1- \sigma_\phi^2/\sigma^2_R$) gives the velocity anisotropy. Often, it is assumed that the velocity dispersion is isotropic, so that $\sigma_\phi = \sigma_R$. However, here, we follow \citet{Anne2008} in which velocity anisotropy ($1- \sigma_\phi^2/\sigma^2_R)$ is given as $ (1-\alpha )/2$, where $\alpha$ is the radial slope of the rotation velocity ($\alpha = \partial lnV_\phi/ \partial ln R$). Therefore, we do not need to limit our formalism to isotropic velocity dispersion. Finally, Equation~\ref{eq:Vadc0} takes the form of:
%but following \citet{Anne2008} we consider that velocity anisotropy ($1- \sigma_\phi^2/\sigma^2_R)$ is   $ (1-\alpha )/2$), where $\alpha$ is the radial slope of the rotation velocity ($\alpha = \partial lnV_\phi/ \partial ln R$). Finally, Equation~\ref{eq:Vadc0} takes the form of:
\begin{equation}
\label{eq:Vadc}
V^{PGC}_{c} = \sqrt{V^2_{\phi} - \sigma^2_R \Big[ \frac{\partial ln\Sigma}{\partial ln R} + \frac{\partial ln\sigma^2_R}{\partial ln R} + \frac{1}{2}(1-\alpha) \Big]} 
\end{equation}
where $V^{PGC}_{c}$ is the pressure gradient corrected circular velocity. Although, we assume velocity anisotropy but this term has a small effect as it is sub-dominant with respect to the combined slope in the density and dispersion. In Figures~\ref{fig:ADC0}, \ref{fig:ADC1} \& \ref{fig:ADC2}, we have shown a few examples of PGC on rising and falling RCs. Notice, BBarolo generated RCs are corrected for pressure support while moment-1 maps are not.

%However, the slope in neither density nor in dispersion can be ignored for high-$z$ galaxies.

%\textbf{From the kinematic modelling of the datacubes (discussed in Section~\ref{sec:Kmodelling}), we have sufficient information about $\Sigma, \sigma_R$ and $V_\phi$ to employ the PGC. Notice, here $\Sigma, \sigma_R$, and $V_\phi$ are derived from $H_\alpha$ gas-kinematics of galaxies. As \citet{Anne2008} have shown that their formalism works well for ionised gas ([OIII]) and stars therefore we assume it is also valid for $H_\alpha$ ionized gas. Moreover, at high-$z$ using KROSS data, only $H_\alpha$ emission is strong enough to trace the kinematics of these galaxies. Therefore we have employed \citet{Anne2008} method to account for the pressure support}. In Figures~\ref{fig:ADC0}, \ref{fig:ADC1} \& \ref{fig:ADC2}, we have shown a few examples of PGC on rising and falling RCs. Notice, BBarolo generated RCs are corrected for pressure support while moment-1 maps are not.

Finally let us remark, \citet{val2007} presented a similar pressure gradient formalism as discussed above. \citet{Dalcanton2010, Burkert2010} and \citet{SW2019} apply the pressure support correction following the previous argument, but assuming the constant dispersion. \citet{read2016} apply a correction following a similar asymmetric drift argument as \citet{val2007}, but also assuming a constant dispersion. Regarding this, we have also explored the pressure support for constant velocity dispersion (see Appendix~\ref{sec:app-adc}).

%%%%%%%%%%%%%%%%%%%%%%%%%%%%%%%%%%%%%%%%%%%%%%%%%%%%%%%%%%%%%%%%%%%%%%%%
\section{RESULTS}
\label{sec:results}
\begin{figure*}
	\begin{center}
		\includegraphics[angle=0,height=7.0truecm,width=18truecm]{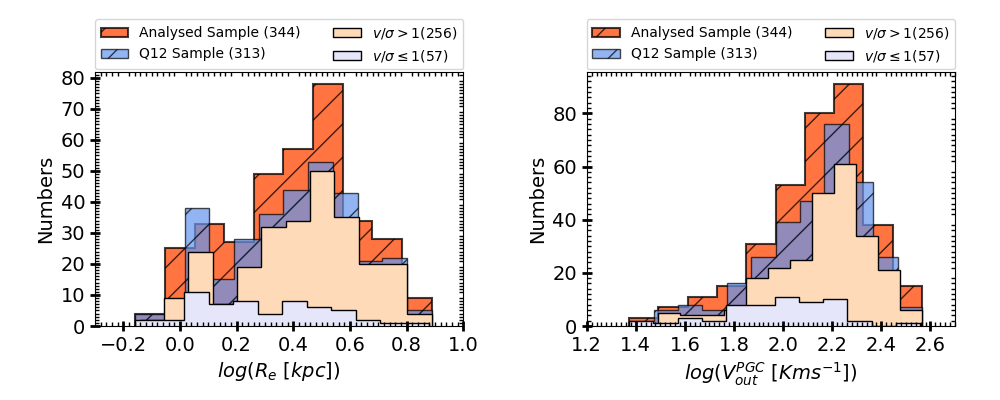}
		\caption{{\em Left:} The effective radii distribution and {\em Right:} the circular velocity distribution of the analysed sample. The colour codes are same in both panel and given as following: the dark orange hatched distribution is the BBarolo analysed sample, blue hatched distribution is Q12 sample (i.e., Quality-1 \& 2 objects only). The peach-puff histogram represents the rotation dominated (i.e., $v/ \sigma >1$) Q12 sample, and the light blue distribution shows to the the dispersion dominated (i.e., $v/ \sigma \leq 1$) Q12 sample. The digits in the bracket of legend represent the total number of the objects in each sub-sample. In the RC analysis, we have used only rotation-dominated Q12 sample.}
		\label{fig:total-data-hist}
	\end{center}
\end{figure*}

In summary, we ran BBarolo on 344 SFGs selected from KROSS survey, having the $H_\alpha$ flux ($F_{H\alpha}>2 \times 10^{-17} \ \mathrm{[erg \ s^{-1}\ cm^{-2}]}$), the inclination ($25^{\circ} \leq \theta_i \leq 75^{\circ}$), the redshift range ($0.75\leq z \leq 1.04$) , and the stellar masses ($8.83 \leq \log (M_{*} \ \mathrm{[M_{\odot}]} ) \leq 11.32$). Thereby, we have 344 beam smearing corrected rotation curves, dispersion curves,  and the corresponding moment maps. After quality assessment of BBarolo outputs, we have selected 314 Quality-1 \& Quality-2 objects; this sample is referred to as the Q12 sample in the analysis (see Section~\ref{sec:Bresults}). In this sample, we have 256 rotation-dominated ($v/ \sigma >1$) and 57 dispersion-dominated ($v/ \sigma \leq 1$) systems, where $v/ \sigma$ is calculated before the PGC. A distribution of re-sampled data is shown in Figure~\ref{fig:total-data-hist}. This sample has objects with the effective radii $-0.16 \leq \log (R_{e} \ \mathrm{[kpc]}) \leq 0.89$ and the circular velocities $1.42 \leq \log \ (V^{PGC}_{out} \ \mathrm{[km \ s^{-1}]}) \leq 2.57$.
%---------------------------------------------------------------------------------
%---------------------------------------------------------------------------------
 		%%%%%%%%%%%%%%%%%%%%%%%%%%%%%%%%%%%%%%%%%%%%%%%%%%%%%%
\subsection{Characteristic Velocity}
\label{sec:char-vel}
\begin{figure}
	\begin{center}
		\includegraphics[width=\columnwidth]{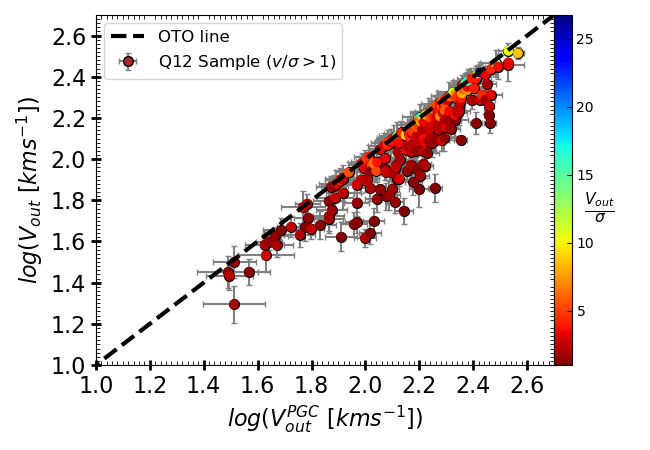}
		\caption{The correlation between rotation velocities before and after pressure gradient correction (PGC), i.e.,$V_{out}$ and $V^{PGC}_{out}$ respectively. The black dashed line shows the one-to-one relation. The objects are color coded by $V_{out}/\sigma$ ($=v/\sigma$, computed before PGC).}
		\label{fig:vout-vadcout}
	\end{center}
\end{figure}

\begin{figure}
	\begin{center}
		\includegraphics[width=\columnwidth]{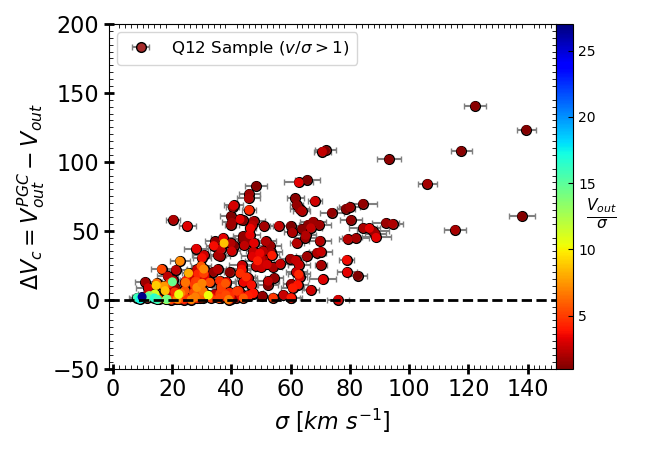}
		\caption{The pressure gradient correction factor: $\Delta V_c = V^{PGC}_{out} - V_{out}$ as a function of velocity dispersion ($\sigma$), color coded for $V_{out}/\sigma$. The corrections are significant in $V_{out}/\sigma \lesssim 5$ galaxies.}
		\label{fig:ADC-cor}
	\end{center}
\end{figure}

For an exponential thin disk, the stellar component of a galaxy follows surface density: $\Sigma_D\left(R \right)=\frac{M_D}{2\pi R_D^2} \exp \left( \frac{-R}{R_D} \right)$ \citep{Freeman}, where $M_D$ is the disk mass and $R_D$ is the disk radius. Under this assumption, one can relate the scale length ($R_e$) determined from the light profile of the galaxy to compute the characteristic radius e.g., the disk length ($R_D=0.59 \ R_e$). In our work, we calculate velocities at three different characteristic radii: \textbf{1)} the optical radius ($R_{opt}=3.2R_D$); \textbf{2)} twice the optical radius ($R_{out}$ = $6.4\ R_D$) and \textbf{3)} $R_{80}$\footnote{$R_{80}$ is the radius which covers the 80\% of the rotation velocity curve} i.e., the radius where the RC is presumably flat (notice that at the different radii, RCs can be rising and falling). The $R_{opt}$ and $R_{out}$ are the photometric measurements (derived from effective radii: $R_e$), whereas, $R_{80}$ is the kinematic measurement. Referring to each radii $R_{opt}$, $R_{out}$ and $R_{80}$, we define the corresponding velocities $V_{opt}$, $V_{out}$ and $V_{80}$ respectively. In this work we have used velocities computed at $R_{out}$. The observed dispersion curves (of ionized gas) are nearly flat, therefore, we have calculated the overall velocity dispersion ($\sigma$) of the galaxy using weighted mean statistics (see: Equation~\ref{eq:WMS}). Note, the characteristic velocities measured from a PG-corrected RCs are referred to as $V^{PGC}_{opt}$, $V^{PGC}_{out}$ and $V^{PGC}_{80}$ respectively. In Figure~\ref{fig:vout-vadcout}, we have shown a comparison of the PGC and the non-PGC circular velocities computed at $R_{out}$ ($V^{PGC}_{out}$ and $V_{out}$ respectively). We can see that if the system does not have sufficient rotation, i.e., if $V_{out}/\sigma \ (= v/\sigma) \lesssim 5$ then the pressure gradient is dominant, which is also noticeable in PG correction factor: $\Delta V_c = V^{PGC}_{out} - V_{out}$ (shown in Figure~\ref{fig:ADC-cor}). On the other hand, if the system is rotation-dominated, then the PGC does not increase the rotation velocity and hence we notice a strong correlation between $V^{PGC}_{out}$ and $V_{out}$ for $v/\sigma >5$. 

Furthermore, we conclude that the pressure gradient is a more dominant effect in the high-$z$ systems than the beam smearing. For a quantitative measurement, we matched RCs of our analysed sample with \citet{H17} RCs, those are not corrected for beam smearing, and calculated the rotation velocity at $3.4 R_D$ in both samples (before and after PGC). We found beam smearing has increased the median rotation velocity of sample by $\sim 10-15 \ \mathrm{km \ s^{-1}}$ (similarly observed by \citealt{HLJ17}), whereas PGC increases the median rotation velocity by more than $30-50$\% relative to the initial value. Although pressure support is dominant than beam smearing, to obtain the `intrinsic shape of RCs', we need to correct for both effects.

 		%%%%%%%%%%%%%%%%%%%%%%%%%%%%%%%%%%%%%%%%%%%%%%%%%%%%%%
\subsection{Co-added Rotation Curves}
\label{sec:CRCs}
Here, we present the five co-added and binned RCs constructed from the 256 individual RCs. We have performed co-addition on both PGC and non-PGC RCs. The technique of co-adding and binning is the following; first, we batch the RCs according to their circular velocity calculated at $R_{out}$, where the dark matter is expected to be dominating. Then we treat each batch of the RCs as a single co-added RC. Second, we bin the galaxies radially per $2.5 \ kpc$ corresponding to the binning scale of BBarolo. For the binning, we have used standard weighted mean statistic given by:
\begin{equation}
\label{eq:WMS}
\bar{X} = \frac{\Sigma_{i=1}^{n}x_i \times w_i}{\Sigma_{i=1}^{n}w_i}
\end{equation}   
where, $ x_i=data; \ w_i=1/error^2; \ \bar{X}=binned \ data$. The errors on binned data are root mean square error, computed as $\delta_{v}^{r_{i}} = \sqrt{\sum([(\delta_{v_{i}}^r)^2 + (\sigma_{v}^{r_{i}})^2])/N}$, where $\delta_{v_{i}}^r$ is the individual error on the velocities per radial bin and $\sigma_v^{r_{i}}$ scatter per radial bin. We are using such a statistical approach because it has plausible advantages on the RC studies e.g., \textbf{a)} it gives us a smooth distribution of RCs ignoring the random fluctuations arising from bad data points, i.e. it allows to enhance the S/N in the data; \textbf{b)} it allows the mass decomposition of similar velocities but having different spatial sampling. This kind of approach in RC studies has been used for decades, pioneered by \citet{PS1991} later developed in several other works \citep{PS1996, PB2000, PS2007, catinella2006, yegorova2011, karukes2017, lapi2018}. To be more precise, we have constructed the five co-added RCs (selected according to the circular velocities at $R_{out}$), where each co-added RC is divided into six radial bins. The statistics, including the number of sources included per radial and velocity bin of the co-added \& binned RCs are tabulated in Table~\ref{tab:RC-stat}. 

%-------------------------
\begin{table}
\centering
\begin{tabular}{|l|l|l|l|l|}
%\begin{tabular}{ llllll}
\hline
Bin Name & N$_v$ & N$_{v,r}$ & $\tilde{R}_{out}$ &  $\tilde{V}^{PGC}_{out}$\\
bin\_$V_{min}$-$V_{max}$ & [dpts] & [dpts] & [$kpc$] &  [$km \ s^{-1}$] \\[1.0ex]
\hline
\hline
bin\_50-100 & 36 & 36, 36, 36, 35, 3  & 12.56  & 80.70 \\
%&  &  &  &  \\	
bin\_100-150 & 65 & 65, 65, 65, 65, 25, 5  & 12.24  & 128.04 \\
%&  &  &  &  \\	
bin\_150-200 & 85 & 85, 85, 85, 85, 45, 25  & 11.79  & 175.51 \\
%&  &  &  &  \\	
bin\_200-250 & 34 & 34, 34, 34, 32, 13, 3  & 13.31  & 220.67 \\
%&  &  &  &  \\	
bin\_250-350 & 24 & 24, 24, 24, 24, 9, 2  & 14.33  & 277.83 \\
\hline
\end{tabular}
\caption{The statistics of the co-added and binned RCs. {\em Col1:} shows the name of each velocity bin, e.g., bin\_a-b, here $a$ is lower velocity limit and $b$ is higher velocity limit. {\em Col2:} N$_v$ gives the number of objects per velocity bin. {\em Col3} N$_{v,r}$ gives the number of objects per radial bin. {\em Col4 \& Col5:} gives the average $R_{out}$ and $V_{out}$ respectively for the each velocity bin. }
\label{tab:RC-stat}
\end{table}

%-------------------------
\begin{figure*}
	\begin{center}
		\includegraphics[angle=0,height=21.0truecm,width=17.5truecm]{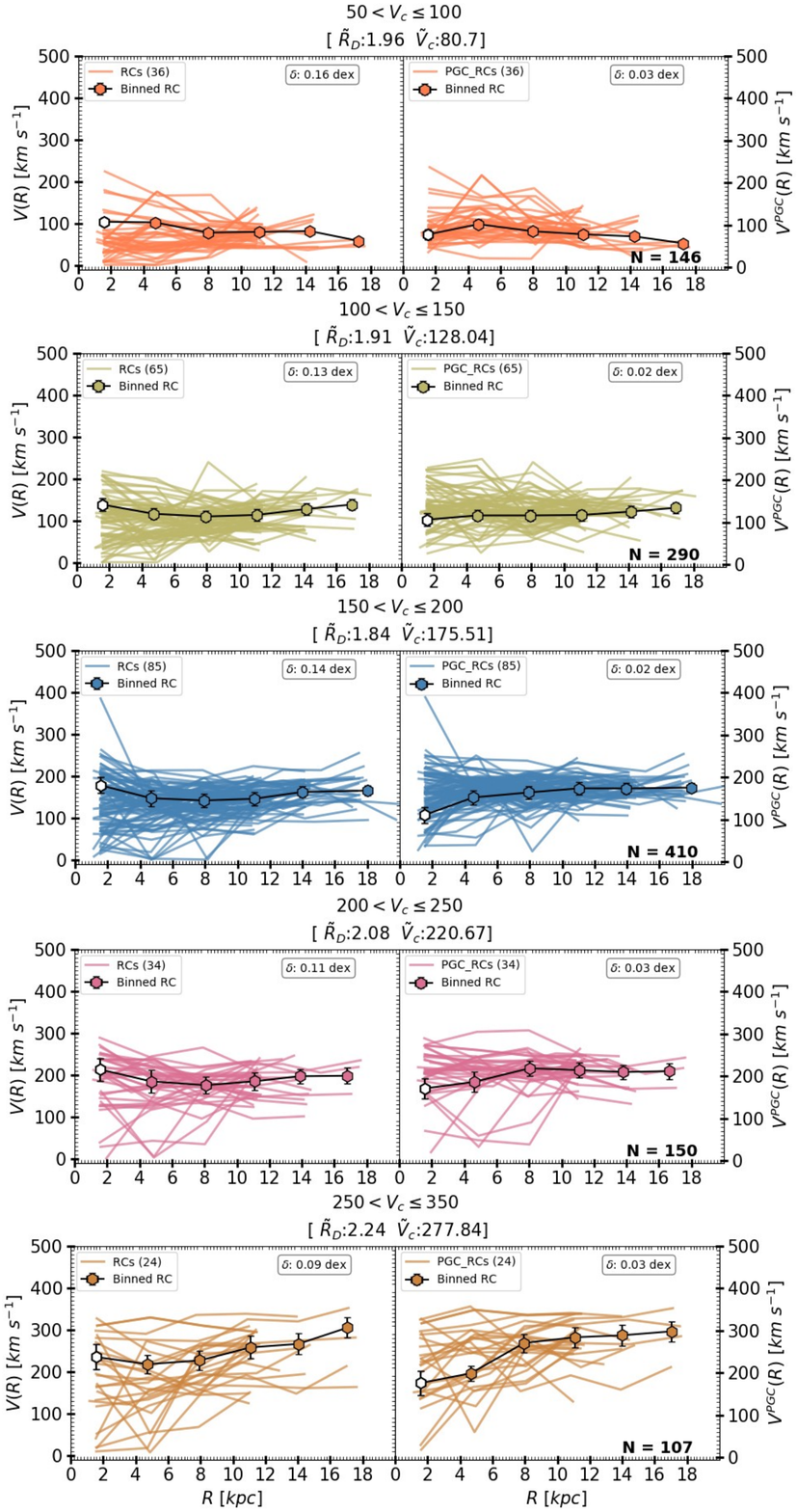}
		\caption{The comparison of non-PGC and PGC co-added RCs (left and right panel respectively). From top to bottom- bin\_50-100: orange colour, bin\_100-150: khaki colour, bin\_150-200: blue colour, bin\_200-250: pink colour and bin\_250-350: dark-orange colour. The color code is same for individual RCs (solid curves/lines) and Co-added RCs (black line/curve connected with hexagons). The median disk radii ($\tilde{R}_D$ in $kpc$) and the median circular velocity ($\tilde{V}_{c}$ in $km \ s^{-1}$) of each bin is displayed in the title of each bin. The digits in the bracket of each legend represent the number of RCs co-added in each bin, and total number of co-added data points per bin are printed in the lower right corner of each plot. The variance in the individual RCs data is given by $\delta$, shown in the rectangular box in the upper-right corner of the each plot. Notice, binned data point of first radial bins is within the resolution limit. Hence, to make it clearer to the reader, we show it by open/white hexagon, and discarded in the further analysis of RCs.}
		\label{fig:RC-ADCRCs}
	\end{center}
\end{figure*}

\begin{figure*}
	\begin{center}		\includegraphics[angle=0,height=8.5truecm,width=16truecm]{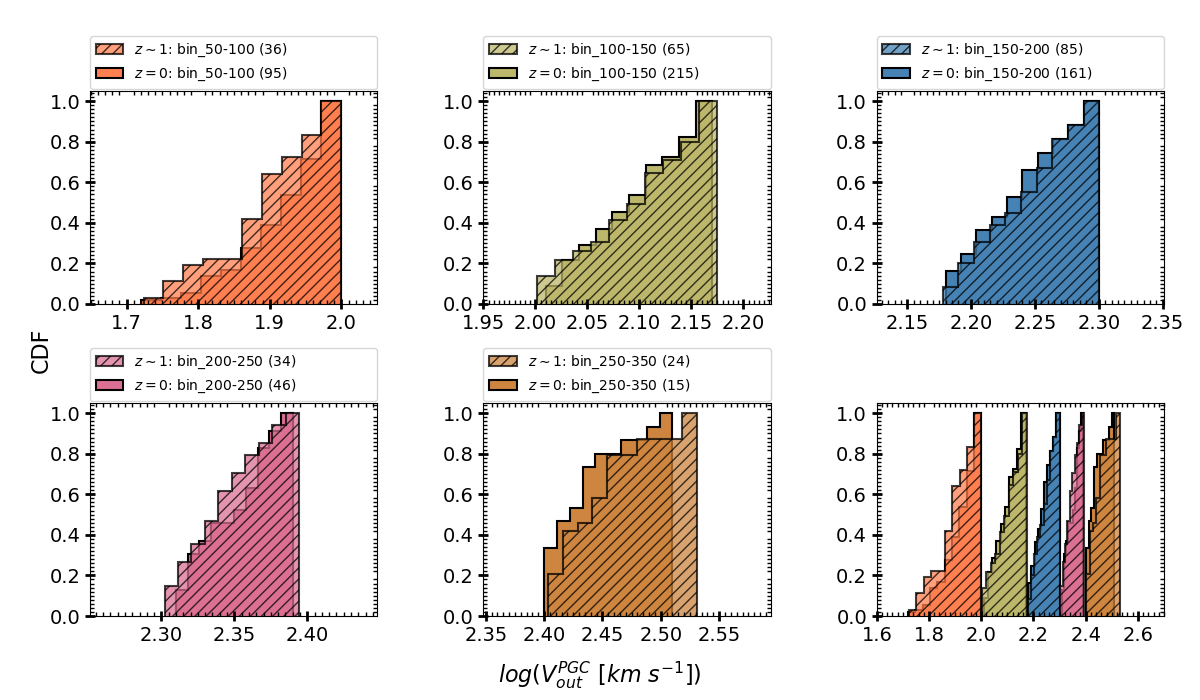}
		\caption{A comparison of cumulative density distribution of circular velocities of $z\sim 1$ (hatched histograms) and $z\approx 0$ (plane histograms) galaxies. The color code of each bin is same as Figure~\ref{fig:RC-ADCRCs}.}
		\label{fig:bin-dist-Vc}
	\end{center}
\end{figure*}

\begin{figure*}
	\begin{center}	\includegraphics[angle=0,height=8.5truecm,width=16truecm]{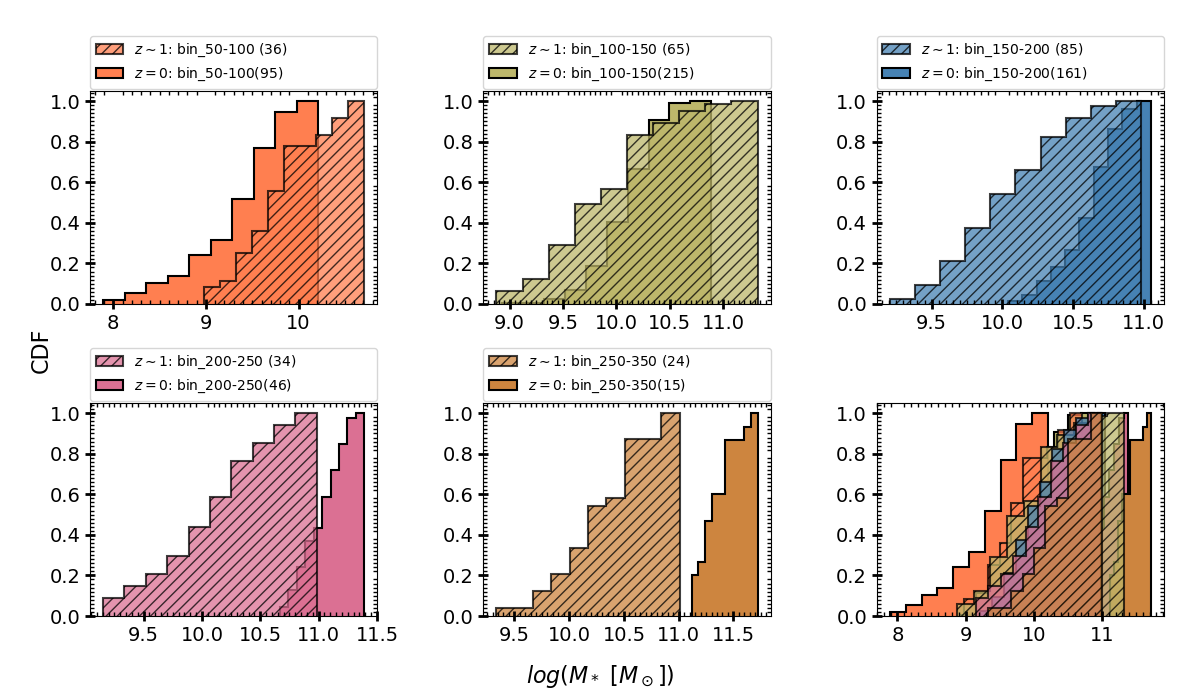}
		\caption{A comparison of cumulative density distribution of stellar masses of $z\sim 1$ (hatched histograms) and $z\approx 0$ (plane histograms) galaxies. The color code of each bin is same as Figure~\ref{fig:RC-ADCRCs}.}
		\label{fig:bin-dist-Ms}
	\end{center}
\end{figure*}

%--------------------------
\begin{figure*}
	\begin{center}
		\includegraphics[angle=0,height=10.0truecm,width=14truecm]{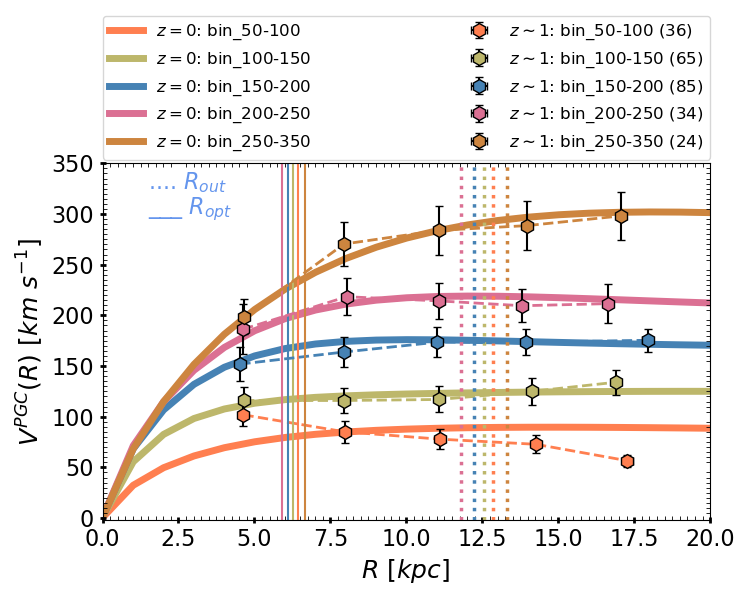}
		\caption{A comparison of $z\sim 1$ co-added RCs of five velocity bins (dashed line connected hexagons) with local RCs (solid curves) of similar velocity bins derived from the URC sample (\citealt{PS1996, PS2007}). The colour code of each velocity bin is the same as Figure~\ref{fig:RC-ADCRCs}. The digits in the bracket of the legend represent the number of RCs co-added in each bin. Dotted and solid vertical lines are showing the $R_{out}$ and $R_{opt}$ respectively for each RC (colour coded same as velocity bins).}
		\label{fig:z0-z1-ADCRCs}
	\end{center}
\end{figure*}

\begin{figure*}
	\begin{center}
		\includegraphics[angle=0,height=9.5truecm,width=13truecm]{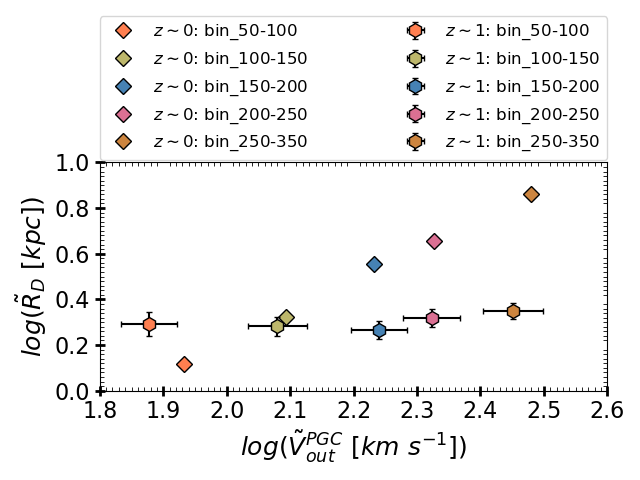}
		\caption{shows the median stellar disk radii ($\tilde{R}_D$) of each velocity bin as a function of their median circular velocities $\tilde{V}^{PGC}_{out}$. The velocity bins are colour coded similarly as Figure~\ref{fig:z0-z1-ADCRCs}. The hexagons with error bars represents the $z\sim 1$ galaxies and diamonds are showing the local galaxies of similar velocity bin.}
		\label{fig:RCa}
	\end{center}
\end{figure*}

%--------------------------

In the Figure~\ref{fig:RC-ADCRCs}, we have shown the co-added \& binned rotation curve before and after PGC. We notice that the some of the individual RCs are subjected to large fluctuations before the PGC. In particular, the individual RCs of the  lower velocity bins ($bin\_50-100 \  \& \ bin\_100-150$) falls to zero (or stay close to zero) at $R>4 \ kpc$ (left panel). This might be interpreted as the galaxies not being in the disk-like configuration. However, it could also be an effect of heavy turbulence (i.e., large pressure gradient), which provides a substantial pressure support, and keeps the disk in kinematic equilibrium without the need for strong rotation.  We suggest that it is most likely the latter case, because after the PGC, we do not see any RC lie close to zero circular velocity (right panel). This effect is also noticeable in characteristic velocity measurements (see Figure~\ref{fig:vout-vadcout}) and PG correction factor (see Figure~\ref{fig:ADC-cor}) of low $v/\sigma (<5)$ galaxies, where the corrections are $\sim 100-150 \ \mathrm{km \ s^{-1}}$. Here, we would like to emphasize the importance of our PGC, which makes use of complete information about the galaxy available in the datacube (namely density, dispersion, and rotation ) and hence is able to correct for the strongly turbulent pressure conditions. The PGC also gives an usual shape to all RCs (as we can see in the right panel of Figure~\ref{fig:RC-ADCRCs}). Moreover, we notice that the variance ($\delta$) in the PG corrected RCs decreased by a factor of $\sim 7$.

We compare our binned RCs of $z\sim 1$ with the Universal Rotation Curves (URCs) of local star-forming disk-type galaxies. The URCs sample is one of the largest (contains 2300 galaxies) and deeply investigated samples, which have been used several times to show the interplay of dark and luminous matter in the local Universe \citep[e.g.,][]{PS1991, PS1996, PB2000, PS2007, catinella2006, yegorova2011, lapi2018}. For comparison, we have used the RCs of the same velocity bins drawn from the URC sample \citep[particularly used in][]{lapi2018}. The distribution of the circular velocities of both samples is shown in Figure~\ref{fig:bin-dist-Vc}. We notice that the circular velocities of both samples trace a similar \textit{total potential} or \textit{total mass} within $R_{out}$. In the Figure~\ref{fig:bin-dist-Ms}, we have shown the distribution of stellar masses of both samples. Notice, for $z\sim 1 $ galaxies stellar masses are derived from SED fitting while for locals (URC sample) they are derived from RC analysis given in \citet{lapi2018}. We found that the stellar masses show a relatively broad distribution for a given velocity bin, and are noticeably lower in comparison with locals (except the lowest velocity bin). Therefore, we can state that both samples are indistinguishable in terms of \textit{total mass} but differ in stellar masses, an outcome which is expected from {\em the empirical galaxy formation and evolution model} \citep[c.f.][and references therein]{Moster2018}. This implies that we are comparing the local SFGs (disk-type systems), with galaxies at $z\sim 1$ those are most likely their progenitors, in which gas is turning into stars keeping the \textit{total mass} constant.

Notice, due to low spatial resolution in the inner region ($R < 2.5 \ kpc$), we do not draw any conclusion within the effective radii. Moreover, due to the lack of knowledge on gaseous components and the fact that stellar disk scale-lengths show no dependency on the circular velocity, we leave the presumably complex process of RC normalization for future work.  Hence, in this work, all the RCs are presented in physical units, i.e., $\mathrm{km \ s^{-1}}$ opposed to $\mathrm{kpc}$. Let us finally remark, the model and techniques employed in this work may be subject to some uncertainties, mostly because  high-$z$ galaxies may not be fully rotation supported disk galaxies. However, this is a problem for any high-$z$ study and can be mitigated only by acquiring better data, e.g., JWST/NIRSpec and ELT/HARMONI as providing more sensitive and higher spatial resolution observations of high-$z$ galaxies.

\section{DISCUSSION}
\label{sec:discussion}
Although we are limited by a few caveats mentioned in previous section, we proceed to interpret our results in the context of $z\approx 0$ star-forming disks. In the Figure~\ref{fig:z0-z1-ADCRCs}, we show the comparison of our binned RCs (PGC applied) of $z\sim 1$ with the URCs of $z\approx 0$. For the purpose of clear and tidy comparison, we do not show the first radial bin of each binned RCs. The last radial bins contains the few data-points  but we keep them in the analysis because they can be informative\footnote{For example, RCs of the lowest velocity bin declines at the last (and second last) point of observation, it could be due to very few data points in the last radial bin. However, pay attention to the clear offset in stellar disk scale length in the lowest bin (see Figure~\ref{fig:RCa}), which apparently indicate high stellar-mass objects relative to $z\approx 0$ galaxies (given that $V_c^{z\sim 1} < V_c^{z\sim 0}$). Therefore, RCs can be declining in the lowest velocity bin. On the other hand, RCs of all the other velocity bins stays flat, even though they contain a few data points in the last radial bins (see Figure~\ref{fig:z0-z1-ADCRCs}).}. We notice that all the $z\sim 1$ RCs coincide with the local RCs from optical radii ($\sim 2R_e$) till the last point of observation. Except the lowest bin ($bin\_50-100$) which shows a gradual decline with increasing radius. In short Figure~\ref{fig:z0-z1-ADCRCs} is an explicit representation of $z\sim 1$ flat RCs. In the Appendix~\ref{fig:RCs-Ms}, we also show the co-added RCs in stellar mass bins. We yet again find the flat RCs with relatively large error in the amplitude. On this basis we conclude that the:
\begin{enumerate}

\item SFGs of $z\sim 1$ manifest flat RCs from optical radii ($\sim 2 R_e$) till last point of observation.

\item Total potential ($\phi(R)$) or total mass within the $\sim 20 \ kpc$ radii of  $z\sim 1$ and $z\approx 0$ SFGs remains the same, i.e., it did not evolve in the past $6.5 \ Gyr$, which suggests that the DM halo of SFGs most likely evolves slowly by accumulating the matter in the outermost regions of the galaxies. These results concord with the theoretical explanation of \citet{Lapi2020}. 
\end{enumerate}

Furthermore, we notice a size-evolution in our sample relative to the local late-type galaxies. For the early and late-type galaxies, size-mass evolution has been already observed and reported in several studies \citep[e.g.,][and references therein ]{VdW2014, Lapi2018a, Tacconi2018}. For an example case, there is an extensive body of work done by \citet{VdW2014} on the size evolution of late-type as a function of redshift, where they bin the galaxies in stellar masses and effective radii for redshift range $0.25\leq z \leq 1.75$. They showed that the size evolution in late-type galaxies could be described as $R_e \propto M_*^{0.22}(1+z)^{-1.48}$. In our work, we are addressing the size evolution in the velocity plane of the $z\sim 1$ galaxies and comparing it with locals, shown in Figure~\ref{fig:RCa}. In particular, we have binned the circular velocities and disk scale-length of galaxies for given velocity bins (same as the RCs in Figure~\ref{fig:z0-z1-ADCRCs}), and plotted the stellar disk radii of each velocity bin ($\tilde{R}_D$) as a function of their circular velocities ($\tilde{V}^{PGC}_{out}$). We notice that stellar distribution does not depend on circular velocity. In particular, the lowest velocity bin is consistent with $z\approx 0$ but at a \textit{smaller} stellar disk-length than $z\sim 1$ on average. On the other hand, we find that with increasing circular velocities, locals are consistent with $z\sim 1$ but contain \textit{larger} stellar disk-lengths. In short, for the high-$z$ galaxies, the stellar disk radii remains constant as a function of the velocity while for locals it increase. This implies that at $z\sim 1$ the stellar-disk component has not yet established a strong co-relation between the total mass or the total potential of the halo within $\sim 20 \ \mathrm{kpc}$ as observed in the locals. It could be a consequence of underlying `complex' astrophysics of the galaxy evolution. For example, most of the baryons are still gaseous at $z\sim 1$, i.e. gaseous-disk dominates. The latter comment has been already mentioned by \citet{Glazebrook2013} and recently deeply investigated and favoured by \citet{Tacconi2018}. Here, from the perspective of the dynamics, we emphasize that the stellar mass distribution evolves over cosmic time in the rotation-dominated system, whereas, a negligible evolution is noticed in the lower end of the velocities. Therefore, we suggest that the evolution of stellar light distribution is circular velocity dependent.

%%%%%%%%%%%%%%%%%%%%%%%%%%%%%%%%%%%%%%%%%%%%%%%%%%%%%%%%%%%%%%%%%%%%%%%%
\section{CONCLUSION}
\label{sec:summary}
In this work, we have analysed the KROSS parent sample \citep{stott2016, H17}. These are IFU based (i.e., 3D data) $H_\alpha$ detected star-forming galaxies at redshift $0.75\leq z \leq 1.04$. On the bases of the $H_\alpha$ flux cut ($F_{H\alpha}>2\times 10^{-17} \ \mathrm{[erg \ s^{-1} \ cm^{-2}]}$), signal-to-noise of the  $H_\alpha$ detection ($S/N>3$) and the inclination angle cut ($25^{\circ}\leq \theta_i \leq 75^{\circ}$), we have selected 344 galaxies for analysis. We have used the $^{3D}$BAROLO (BBarolo) for kinematic modelling of these galaxies, which is also capable of extracting the corresponding rotation curves. The main advantage of this tool is that, it incorporate the beam smearing corrections simultaneously with kinematic modelling in 3D-space. Thus it allows us to determine the unbiased rotation velocity and intrinsic velocity dispersion even in the low spatial resolution data. Moreover, we have corrected the RCs for the pressure support applying the pressure gradient correction (PGC), which is discussed in detail in Section~\ref{sec:ADC}. A 3-fold approach of deriving RCs (3D-kinematic modelling + beam smearing corrections in 3D-space + pressure support corrections), delivers the true intrinsic shape of the RCs (see Figure~\ref{fig:RC-ADCRCs} right panel and Figure~\ref{fig:z0-z1-ADCRCs}).

We have analysed the 256 rotation dominated galaxies. This sample covers the redshift range $0.75\leq z \leq 1.04$, the effective radii $-0.16 \leq \log \left(R_{e} \ \left[\mathrm{kpc}\right] \right) \leq 0.89$ , the circular velocities $1.42 \leq \log \left(V^{PGC}_{out} \ \left[\mathrm{km \ s^{-1}}\right] \right) \leq 2.57$, and the stellar masses $8.83 \leq \log \left(M_{*} \ \left[\mathrm{M_{\odot}}\right] \right) \leq 11.32$. Using the technique of \citet{PS1996}, we have constructed the co-added and binned RCs of the five velocity bins out of 256 individual RCs and compared them with the RCs of local star-forming disk-type galaxies of same velocity bins (see: Section~\ref{sec:discussion} and Figure~\ref{fig:z0-z1-ADCRCs}). The main findings of this work are the following:

\begin{itemize}
	\item The pressure gradient is more dominant effect in high-$z$ galaxies than the beam smearing. It corrects the median rotation velocity more than the 50\%, especially  in the galaxies with $v/ \sigma < 5$.
	
	\item A statistically robust method of co-adding and binning RCs shows that the $z\sim 1$ outer RCs are very similar to the outer RCs of the local star-forming galaxies (see Figure~\ref{fig:z0-z1-ADCRCs}) where dark matter dominates and flattens the RCs.
	
	\item We have noticed a significant evolution in the disk scale length over past $ 6.5 \ Gyrs$ (see Figure~\ref{fig:RCa}).

\end{itemize}

On the bases of above outcomes, we conclude that the \textit{Total Matter} placed in a galactic halo within $\sim 20 \ kpc$ radius at $z\sim 1$ remains same as $z\approx 0$. At the same time, stellar mass distribution (i.e., stellar disk) evolved over cosmic time (in past $6.5 \ Gyrs$). This suggests a prolonged evolution of the SFGs (late-type systems).

%%%%%%%%%%%%%%%%%%%%%%%%%%%%%%%%%%%%%%%%%%%%%%%%%%%%%%%%%%%%%%%%%%%%
\subsection*{Acknowledgements}
We thank the anonymous referees for their constructive comments and suggestions, which have significantly improved the quality of the manuscript. We thank A. Tiley for passing us the SED driven stellar masses of KROSS sample. G.S. thanks Enrico M. Di Teodoro, M. Petac, and Luigi Danese for their fruitful discussion and immense support in the entire period of this work. GvdV acknowledges funding from the European Research Council (ERC) under the European Union's Horizon 2020 research and innovation programme under grant agreement No 724857 (Consolidator Grant ArcheoDyn).

%%%%%%%%%%%%%%%%%%%%%%%%%%%%%%%%%%%%%%%%%%%%%%%%%%%%%%%%%%%%%%%%%%%%%%%%%%%%%
\subsection*{Data Availability}
In this work, we make use of KROSS data, which is publicly available at \href{http://astro.dur.ac.uk/KROSS/data.html}{KROSS website}. A catalog of 344 KROSS star-forming galaxies used in the work is available at \href{https://doi.org/10.7910/DVN/MHRG4O}{Gauri Sharma Dataverse}. 
%%%%%%%%%%%%%%%%%%%%%%%%%%%%%%%%%%%%%%%%%%%%%%%%%%%%%%%%%%%%%%%%%%%%%%%%%%%%%
\bibliographystyle{mnras}
\bibliography{gsharma2020.bib}

%%%%%%%%%%%%%%%%%%%%%%%%%%%%%%%%%%%%%%%%%%%%%%%%%%%%%%%%%%%%%%%%%%%%%%%%%%%%%
 
\appendix

\section{Kinematic and Photometric Position Angles}
\label{sec:kin-phot-barolo}

\begin{figure*}
	\begin{center}
		\includegraphics[width=\columnwidth]{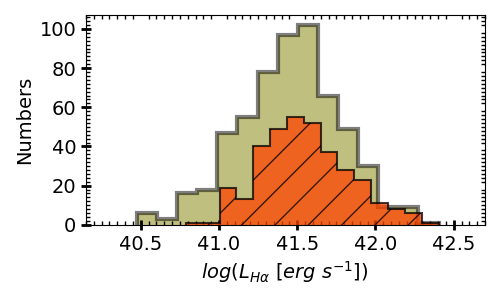}
		\includegraphics[width=\columnwidth]{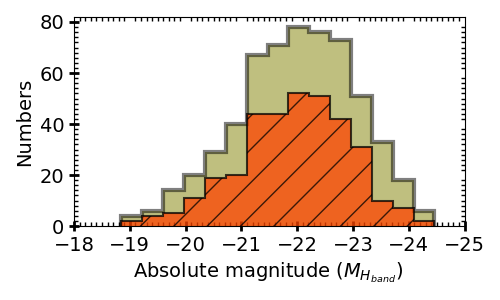}
		\caption{Figure\ref{fig:KROSS-hist} continued. {\em Left} $H_\alpha$ luminosity, {\em Right} $H$-band absolute magnitude.}
		\label{fig:extrahist}
	\end{center}
\end{figure*}

\begin{figure*}
	\begin{centering}
		\includegraphics[angle=0,height=4.5truecm,width=18truecm]{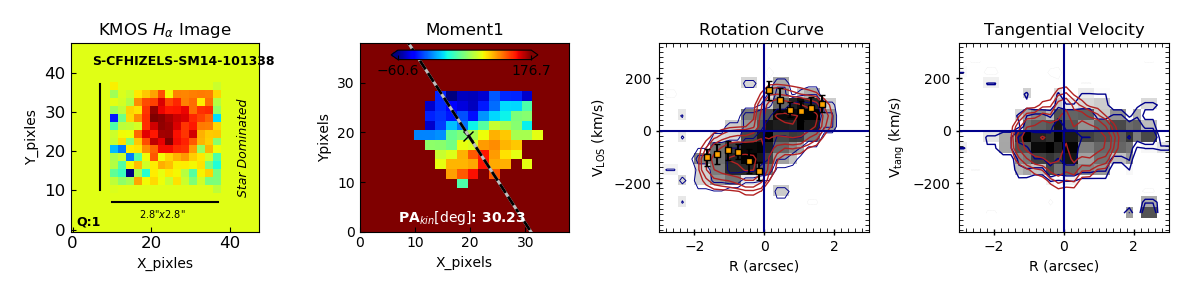}		
		\includegraphics[angle=0,height=4.0truecm,width=18truecm]{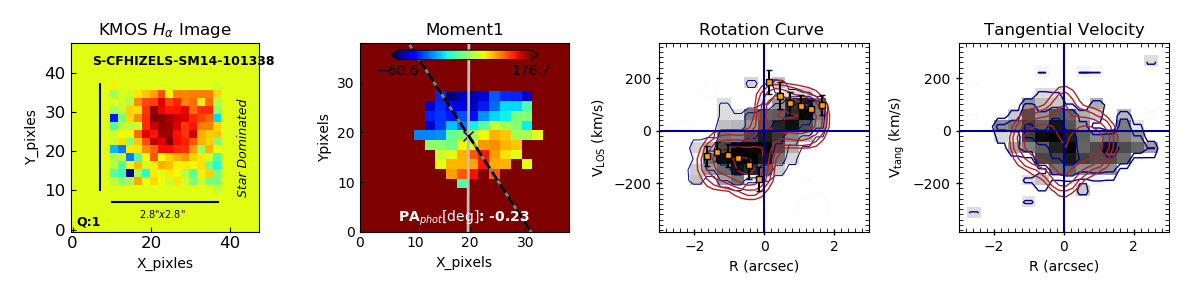}
		\caption{{\em Upper Panel:} Kinematic modelling using $PA_{kin}$. {\em Lower Panel:} Kinematic modelling performed using $PA_{phot}$. {\em COL 1:} Integrated $H_\alpha$-image from 3D datacube; {\em COL 2:} Rotation velocity map, where $PA_{kin}$ is shown by dotted dashed black-grey line, and $PA_{phot}$ is shown by solid grey line; {\em COL 3:} LOS rotation velocity curve; {\em COL 4:} Tangential velocity profile. In the PV-diagrams, the black shaded area with blue contour represents the data, the red contour is model, and orange squares with error bars are best-fit velocity measurements. }
		\label{fig:PA_phot_kin_model}	
	\end{centering}
\end{figure*}

If the position angle is correct, then the kinematic modelling allows to extract the PV-diagram symmetric on the $x$ and $y$-axis, but if $PA$ is wrong, then an asymmetry is spotted in PV-diagram (see figure~\ref{fig:PA_phot_kin_model}). For nearly $44\%$ objects $PA_{phot}$ do not coincide with $PA_{kin}$. Therefore we decided to free the $PA$ parameter in BBarolo-run particularly for these objects. The BBarolo estimated $PA$ are referred to as $PA_{kin}$ in the analysis and flagged in the given catalog (Table~\ref{tab:cat}). An example of kinematic modelling using $PA_{phot}$  (when it is wrong) and $PA_{kin}$ is shown in Figure~\ref{fig:PA_phot_kin_model}. {\em Upper panel } shows the kinematics and rotation curve derived using  $PA_{kin}$, whereas {\em Lower panel } shows the kinematics and rotation curve derived from fixed $PA_{phot}$. We can see that PV-diagrams drove from fixed $PA_{phot}$ are asymmetric around $x$ and $y$-axis, while this problem resolved if the $PA$ is set to free in BBarolo run. We are not keeping  $PA$ free for all objects because BBarolo documentation suggests to `keep the number of free parameter low, specifically in low-resolution data'.

\begin{figure}
	\begin{center}
		\includegraphics[width=\columnwidth]{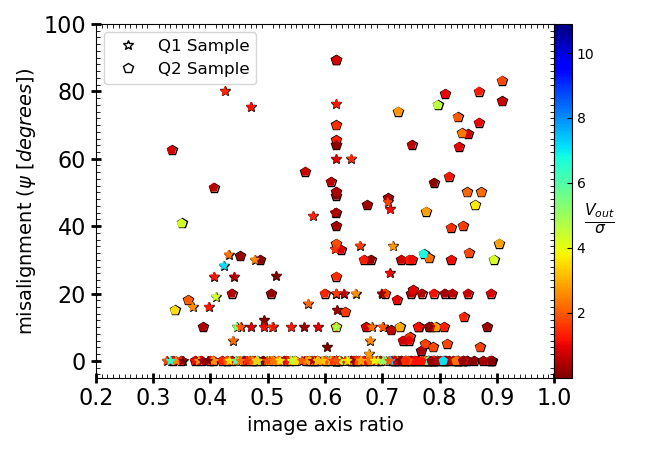}

		\caption{The difference between the photometric and kinematic position angles ($\psi$) as a function of broad-band image axis ratio. The stars and hexagons represents the Q1 and Q2 sample respectively, color coded for $v/\sigma$. Nearly $3.5\%$ Q1-sample and $13.3\%$ Q2-sample have misalignment more than $\psi >30^\circ$. Overall $ 83\%$ Q12-sample have good position angles measures (i.e., $PA_{phot} \simeq PA_{kin}$) which places confidence on our measurements.}
		\label{fig:misalignment}
	\end{center}
\end{figure}

\subsection{Photometric \& kinematic major axes}
\label{sec:misalignment}
Here, we discuss the misaligned kinematic and photometric (morphological) major axes. As we mentioned in the Section~\ref{sec:InitialParams},  nearly $44\%$ of analysed sample have $PA_{phot} \neq PA_{kin}$. This discrepancy in the position angles can affect our results in two ways: \textit{(1)} uncertain inclination angle because inclination is derived using photometric axis ratio, which is used in computing $PA_{phot}$  too (see Section~\ref{sec:sample}) and \textit{(2)} presumably misaligned systems are not in disk configuration, which is worrisome. Therefore, we quantitatively explored the misalignment in the Q12 sample. Following \citet{Wisnioski2015} we define the misalignment between photometric and kinematic position angle as:
\begin{equation}
\label{eq:misalignment}
\sin \psi = |\sin (PA_{phot}- PA_{kin})|
\end{equation}
where $\psi$ gives the misalignment and lies between $0^\circ$ to $90^\circ$. If $\psi >30^\circ$ than morphologies are considered irregular.  Figure~\ref{fig:misalignment}, shows the $\psi$ as a function of image axis ratio, where axis-ratios are derived using broad-band images (for details see \citealt{H17}). We notice, $\sim 40\%$ objects of Q12-sample are misaligned, in which only $\sim 17\%$ objects have $\psi>30^\circ$ (which includes $\sim 3.5 \%$ Q1-sample and $\sim 13.3\%$ Q2-sample). We notice low $v/\sigma$ galaxies are highly misaligned, most likely due to lack of well defined kinematic axis. Despite the different kinematic modelling approaches, our misalignment results are similar to \citet{H17}. Moreover, majority ($\sim 83\%$) of our sample used in RC analysis (Q12 sample), have $PA_{phot} \simeq PA_{kin}$ which places confidence on our measurements. Besides it, we provide a statistical study of RCs which encompass any uncertainties caused by misalignment.

Furthermore, our main results (see Figure~\ref{fig:z0-z1-ADCRCs}) show that the co-added RCs of our sample ($z\sim 1$) coincides with the RCs of local disk galaxies with a precision of more than 95\%. Therefore, we infer that the majority ($\gtrsim 83\%$) of our galaxies are in disk-like configuration. 

%%%%%%%%%%%%%%%%%%%%%%%%%%%%%%%%%%%%%%%%%%%%%%%%%%%%%%%%%%%%%%%%%%%
\section{Examples of Pressure Gradient Corrections}
\begin{figure*}
	\begin{center}
		\includegraphics[angle=0,height=7.5truecm,width=16truecm]{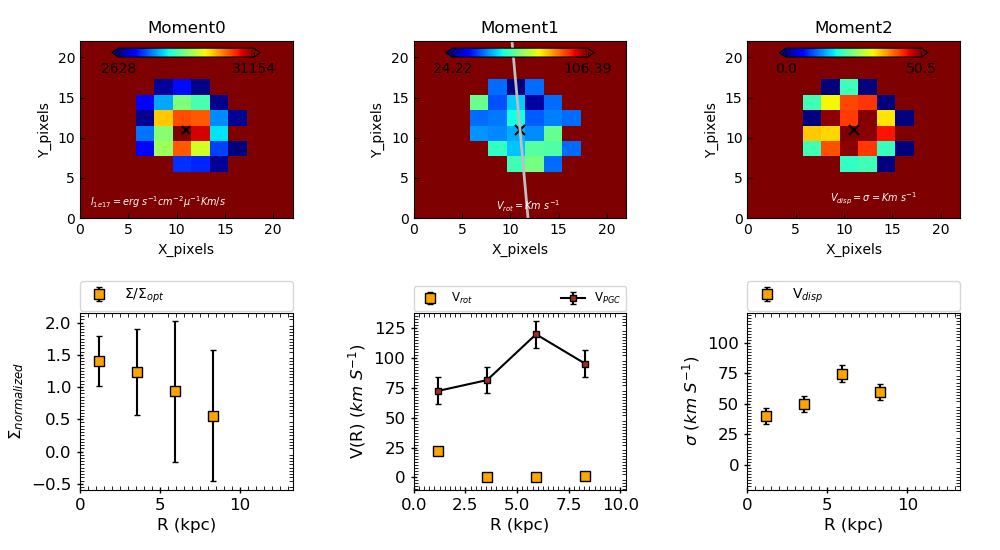}
		\caption{An example of PGC. {\em Upper panel:} left-to-right, zeroth, first, and second moment maps. {\em Lower panel:} left-to-right, the surface brightness, the rotation velocity, and the velocity dispersion. In the moment-1 map, the grey line is the photometric position angle and the black cross are the galactic kinematic center ($x^k_c,y^k_c$). In the surface brightness profile and the P-V diagrams, the orange squares with error bars are best fit data. In the rotation curve profile, brown squares connected with the black line is PG corrected rotation velocity profile.}
		\label{fig:ADC0}
	\end{center}
\end{figure*}

\begin{figure*}
	\begin{center}
		\includegraphics[angle=0,height=7.5truecm,width=16truecm]{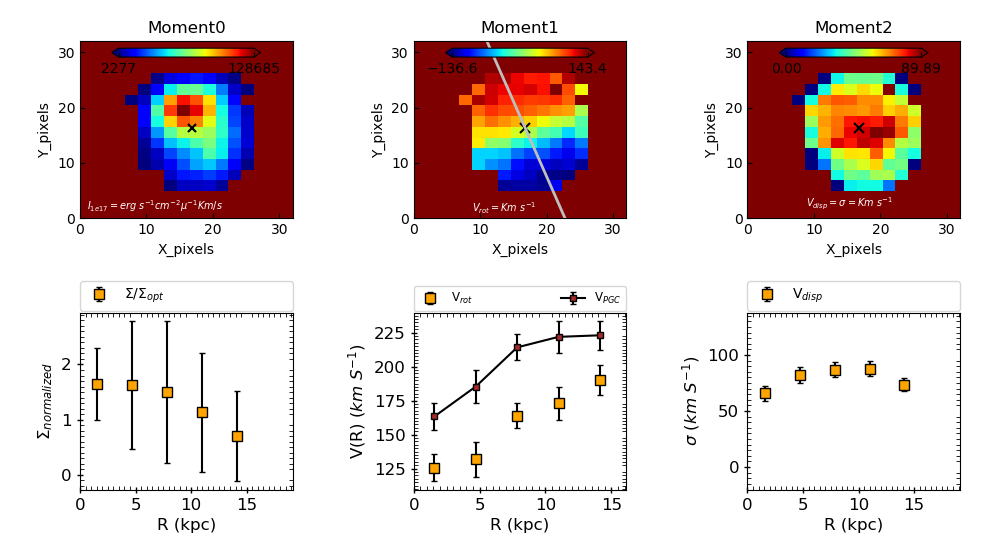}
		\includegraphics[angle=0,height=7.5truecm,width=16truecm]{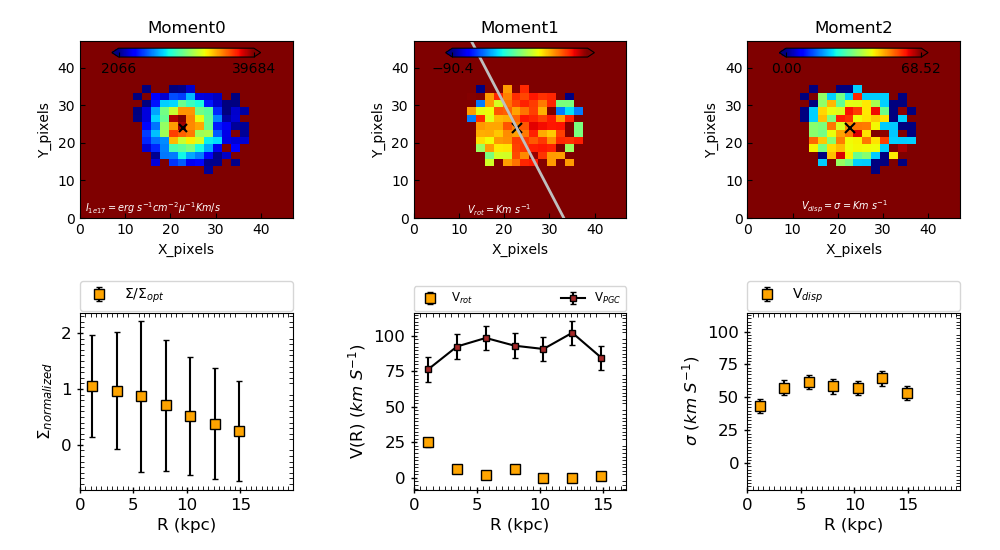}
		\includegraphics[angle=0,height=7.5truecm,width=16truecm]{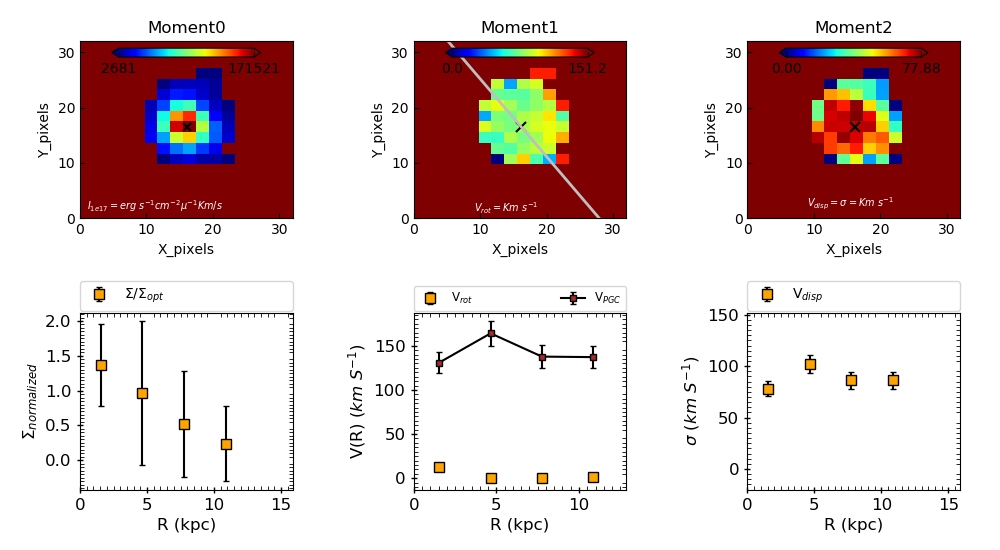}
		\caption{PGC examples continue.. }
		\label{fig:ADC1}
	\end{center}
\end{figure*}

\begin{figure*}
	\begin{center}
		\includegraphics[angle=0,height=7.5truecm,width=16truecm]{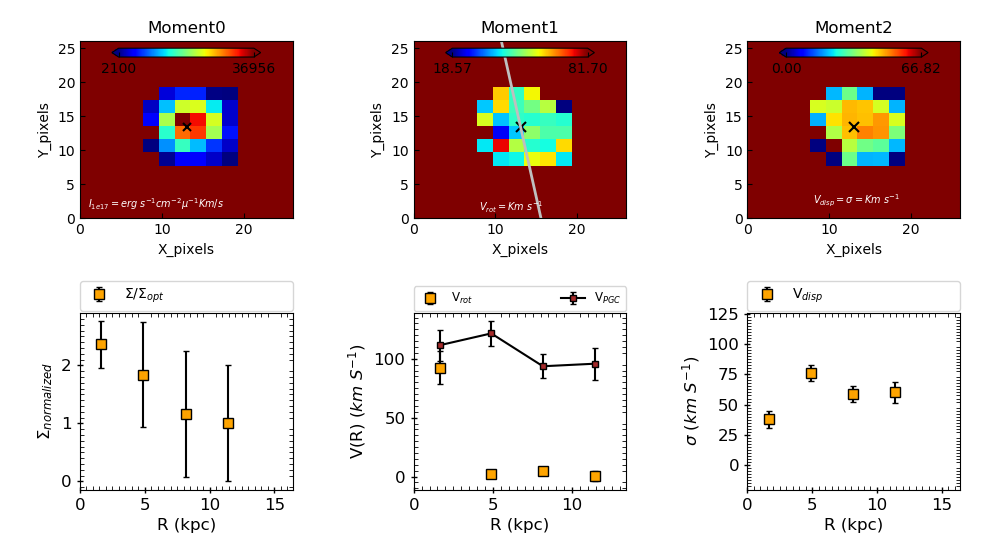}
		\includegraphics[angle=0,height=7.5truecm,width=16truecm]{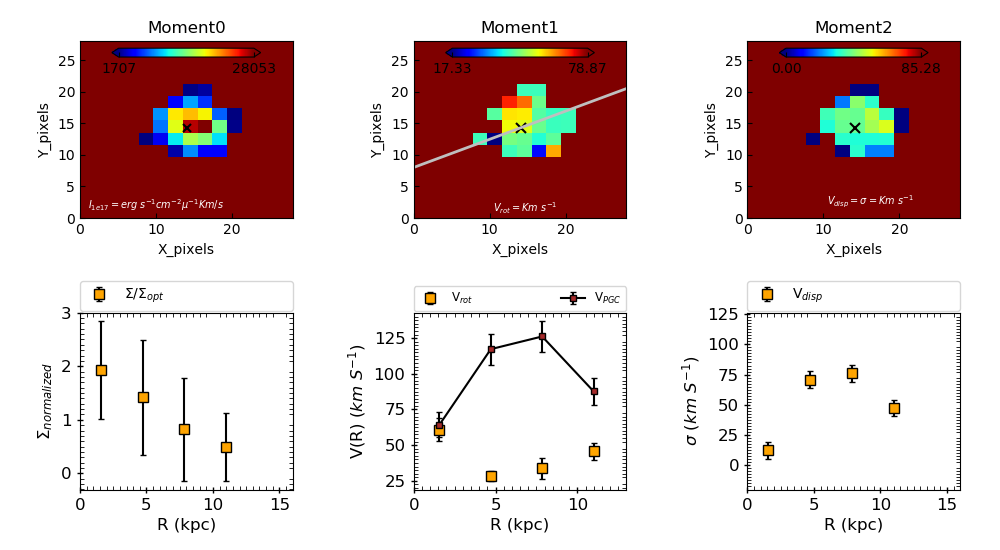}
		\includegraphics[angle=0,height=7.5truecm,width=16truecm]{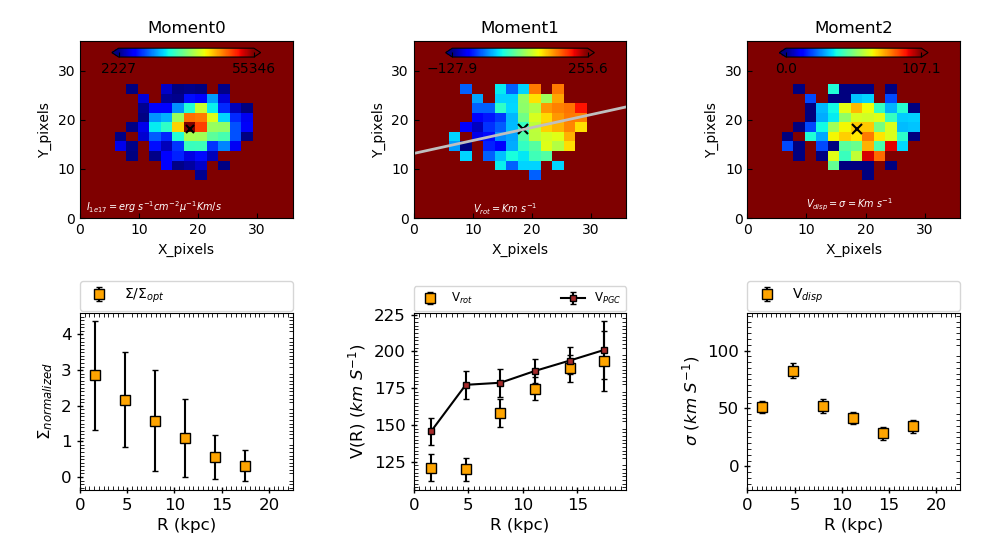}				
		\caption{PGC examples continue..}
		\label{fig:ADC2}
	\end{center}
\end{figure*}

Some Examples of Pressure Gradient Correction (PGC) on the BBarolo generated rotation curves are shown in the Figures~\ref{fig:ADC0}, \ref{fig:ADC1} \& \ref{fig:ADC2}.  

\section{Pressure Gradient Corrections with Constant $\sigma$}
\label{sec:app-adc}
\begin{figure*}
    \begin{center}
        \includegraphics[angle=0,height=9.5truecm,width=18truecm]{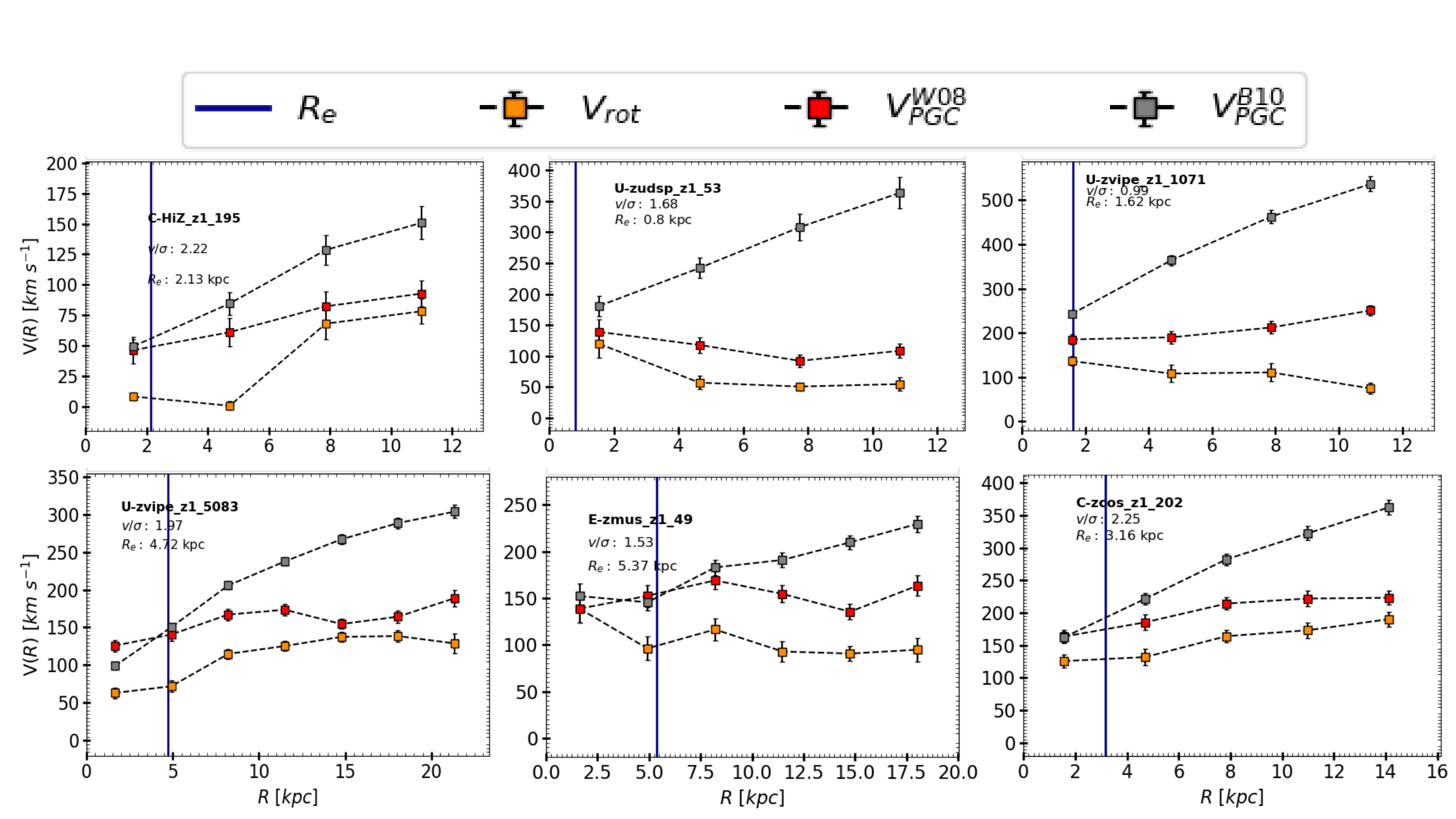}       
        \caption{The Rotation curves before and after PGC using  \citet{Anne2008} and \citet{Burkert2010} method. The yellow square represents the intrinsic RC without PGC (denoted by $V_{rot}$). The red and gray square shows the PG corrected RCs using \citet{Anne2008} and \citet{Burkert2010} method, denoted by  $V^{W08}_{PGC}$ and $V_{PGC}^{B10}$ respectively. The vertical blue line shows the effective radius of the galaxy. The name, $v/\sigma$ and $R_e$ of galaxies are printed at upper-left corner of each plot.}
        \label{fig:RC-ADC}
    \end{center}
\end{figure*}

\begin{figure*}
    \begin{center}
        \includegraphics[angle=0,height=6.0truecm,width=18truecm]{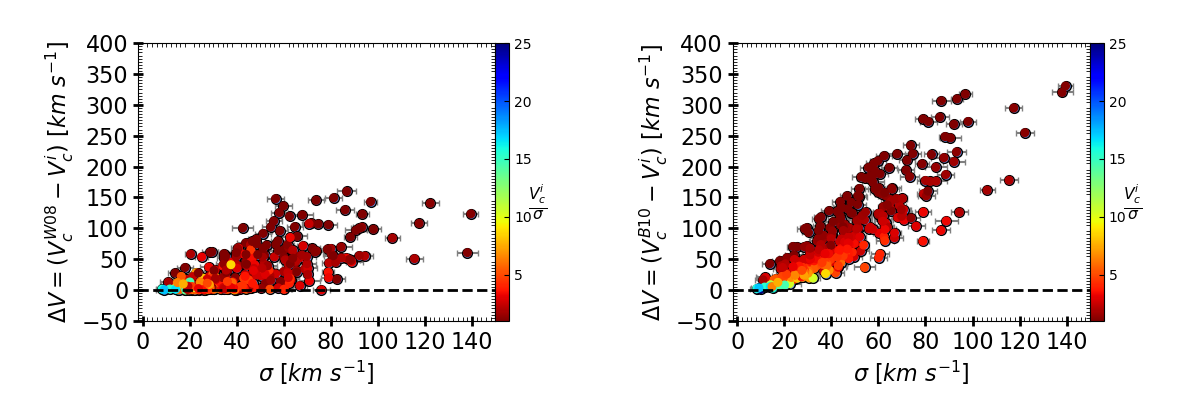}
        \caption{The comparison of correction factor ($\Delta V = V^{PGC}_{c} - V^i_{c} $) on rotation velocities before and after the PGC, using \citet{Anne2008} and \citet{Burkert2010} method (left and right panel respectively). The correction factor is plotted as a function of velocity dispersion and data-points are color coded for $V_c^i/\sigma$ (i.e., $v/\sigma$). The notation of circular velocities are given as follows, $V_c^{W08}$ and $V_c^{B10}$ denotes the PGC using W08 and B10 method respectively. Whereas $V^i_c$ is the inclination corrected rotation velocity without PGC, computed at $R_{out}$. We notice, pressure correction using B10 method are on an average a factor of 2 higher than W08 method, which is also shown by \citet{kretschmer2020}. }
        \label{fig:V-ADC}
    \end{center}
\end{figure*}

We compare our PGC following the \citet[][hereafter W08]{Anne2008} with the often adopted PGC following \citet[][hereafter B10]{Burkert2010} given by:
\begin{equation}
\label{eq:ADC-stat}
V^2_{c}(r) = V^2_{0}(r) + 2\sigma_0^2 \frac{r}{R_D}
\end{equation}
where, $V_{0}$ is the inclination corrected rotation velocity when the pressure gradient is negligible, and the velocity dispersion ($\sigma_0$) is isotropic and radially constant. In Equation[\ref{eq:ADC-stat}] second term gives the pressure term. We have noticed a significant difference in both methods, shown in figure~\ref{fig:RC-ADC}. In B10 method RC are successively rising, and corrections are always applied even when they are not required. This is due to the constant $\sigma$ and $r/R_D$ factor, which imposes the circular velocity to increase as a function of radius. On the other hand, W08 method uses the full information available in the datacubes, including the radially varying surface-brightness, rotation velocity and velocity dispersion.  This avoids the un-necessary radially growing circular velocity, as well as, the corrections are only applied if needed. Furthermore, In figure~\ref{fig:V-ADC}, we have shown the amount of pressure support correction $\Delta V = V^{PGC}_{c} - V^i_{c}$ (computed at $R_{out}$) in both cases as a function of velocity dispersion. We can see that the corrections are twice higher in B10 method than W08, and unrealistically high in low $v/\sigma$ galaxies. 

Recently, based on high-resolution zoom-in cosmological simulation, \citet{kretschmer2020} compared the pressure term of jeans/hydrostatic equilibrium \citep{Anne2008} given by:
 \[ \alpha_{true} = - \frac{\partial \ ln \ \rho \ \sigma^2_R}{\partial \ ln \ R} + (1-\frac{\sigma^2_\phi}{\sigma^2_R}) + \Delta Q \]
where, all the symbols have their usual meaning (as given in Section~\ref{sec:ADC}) and $\Delta Q$ is non-spherical potential term, which we neglect in our work. On the other hand, pressure term in Self-gravitating Exponential Disc \citep{Burkert2010}, given by:
\[\alpha_{self}=2 \frac{r}{0.59 \ R_e} \] 
 
\citet{kretschmer2020} concluded that 1) $\alpha_{self} = 2 \ \alpha_{true}$, 2) pressure correction from non-isotropic and non-constant velocity dispersion are not negligible for gas within $R=R_{e,gas}$, and 3) pressure support correction derived from self-gravitating disk assuming constant velocity dispersion is not valid for high-$z$ galaxies. In Figure[\ref{fig:V-ADC}], we clearly verify their findings. Moreover, since we do not know the exact distribution of gas in high-$z$. Therefore, we encouraged to avoid any prior assumption on gas component, whether it is related to turbulent pressure or effective radii of gaseous disk ($R_{e,gas}$).

    %%%%%%%%%%%%%%%%%%%%%%%%%%%%%%%%%%%%%%%%%%%%%%%%%%%%%%%
\section{Co-added RCs of Stellar Mass Bins}
\begin{figure*}
	\begin{center}	\includegraphics[angle=0,height=10truecm,width=16truecm]{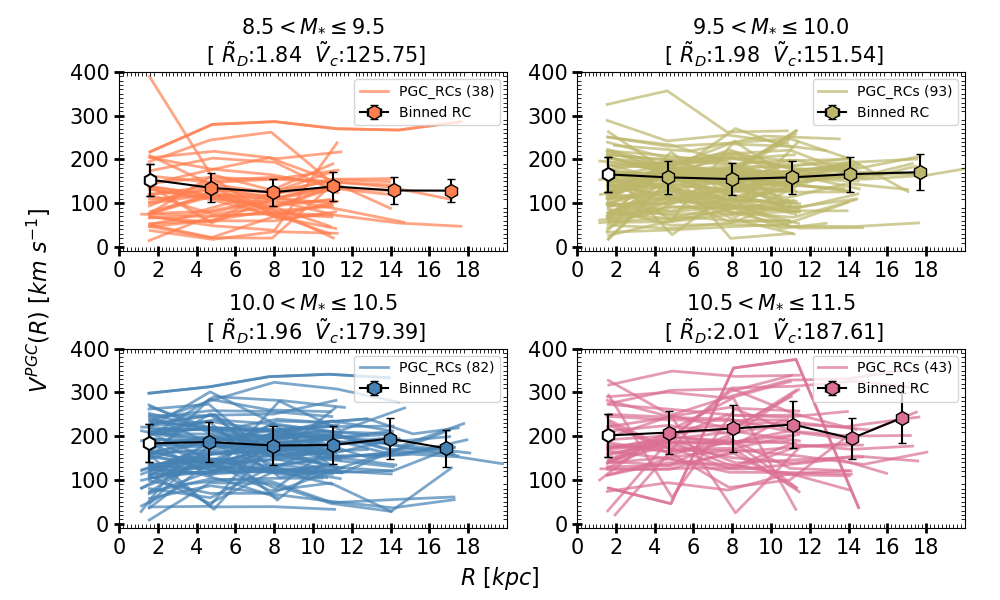}
		\caption{The co-added and binned RCs in the stellar mass bins. The color code of $M_*$ bins is the following, the orange: bin\_8.5-9.5, green: bin\_9.5-10, blue: bin\_10.0-10.5, and pink: bin\_10.5-11.5. The solid lines represents the individual RCs and hexagon connected black curve shows the binned RCs. The procedure of co-adding and binning is same as given in the Section~\ref{sec:CRCs}}
		\label{fig:RCs-Ms}
	\end{center}
\end{figure*}

In Figure~\ref{fig:RCs-Ms}, we show the co-added \& binned RCs of four stellar mass bins, namely  bin\_8.5-9.5, bin\_9.5-10, bin\_10.0-10.5, and bin\_10.5-11.5. We yet again find the flat RCs with relatively higher error in the amplitude.

%%%%%%%%%%%%%%%%%%%%%%%%%%%%%%%%%%%%%%%%%%%%%%%%%%%%%%%%%%%%%%%
\section{Catalogue}
With this paper we release a catalog of raw values (adopted from \citealt{H17}) and derived values for the 344 KROSS star-forming galaxies used in the work. In Table~\ref{tab:cat} we describe the columns of the catalog released with this paper. The full version of Figure~\ref{fig:barolo-output} and a machine readable table is attached in the external appendix.

\begin{table*}
	{\footnotesize
		{\centerline {\sc Descriptions of Columns}}
		\begin{tabularx}{\textwidth}{lccl}
			\hline
			Number & Name & Units & Description \\
			\hline
			\hline
			1 & KID & & KROSS ID. \citep[][hereafter Ref: H17]{H17}.\\
			2 & Name && Object Name (Ref: H17).\\
			3 &Quality && Quality flag for the data (see Section~\ref{sec:Bresults}):\\
			&&&{\em Quality 1}: H$\alpha$ detected, best moment map and RCs.\\
			&&&{\em Quality 2}: H$\alpha$ detected, reasonably good moment map and RCs.\\ 
			&&&{\em Quality 3}: H$\alpha$ detected but bad moment maps and RCs (discarded from the analysis).\\	
		
			4 &Flag\_$PA$ && Position angle flag for the data:\\
			&&&{\em kin}: Position angle derived from BBarolo Kinematic modelling.\\ 
			&&&{\em phot}: Photometric position angle (Ref: H17).\\	
						
			5 & RA && Right Ascension [J2000] (Ref: H17)\\
			
			6 & DEC && Declination [J2000] (Ref: H17).\\
			
			7 & z && Redshift from H$\alpha$ (Ref: H17).\\
			
			8 &$PA$ & degrees &Positional angle ($PA$).\\
			
			9 &INC & degrees &Inclination angle (Ref: H17).\\
			
			10 & M\_H && Absolute $H$-band magnitude (Ref: H17)\\
			
			11 & z\_AB && $z$-band AB-magnitude (Ref: H17).\\
			
			12 & K\_AB && $K$-band AB-magnitude (Ref: H17).\\
			
			13--14 & Re, Re\_err &$\log\left(kpc \right)$& Deconvolved continuum half-light radii, $R_{1/2}$, from the
			image and error (Ref: H17).\\
			
			15--16 & Vout, Vout\_err &$\log\left(km \ s^{-1} \right)$& Rotation velocity computed at $6.4 \ R_D$ ($R_D=0.59 \ R_e$) and error.\\
			
			17--18 & VPGC\_out, VPGC\_out\_err &$\log\left(km \ s^{-1} \right)$& PGC corrected Rotation velocity computed at $6.4 \ R_D$ ($R_D=0.59 \ R_e$) and error.\\
			
			19--20 & sigma, sigma\_err &$\log\left(km \ s^{-1} \right)$&  Weighted mean from intrinsic velocity dispersion profile and error.\\
			
			21--22 & L\_Ha, L\_Ha\_err &$\log\left(erg \ sec^{-1} \right)$& Observed H$\alpha$ luminosity and error (Ref: H17).\\
			
			23 & F\_Ha &$\log\left(erg \ sec^{-1} \ cm^{-2} \right)$& Observed aperture H$\alpha$ flux (Ref: H17).\\
			
			\hline
		\end{tabularx}
	}
	\caption{\label{tab:cat}
		Details of the columns provided in catalog associated with this paper. 
	}
\end{table*}

\bsp
\label{lastpage}
\end{document}